\documentclass[iop]{emulateapj}
\usepackage{amsmath,amssymb,latexsym, graphicx}
\usepackage{epsfig}
\usepackage{booktabs}
\usepackage{afterpage,lscape}
\usepackage[dvips]{color}

\begin{document}

\title{Phenomenology of Reverse-Shock Emission in the Optical Afterglows of Gamma Ray Bursts}

\author{J. Japelj$^{1}$, D. Kopa\v{c}$^{1}$, S. Kobayashi$^{2}$, R. Harrison$^{2}$, C. Guidorzi$^{3}$, F. J. Virgili$^{2}$, C. G. Mundell$^{2}$, A. Melandri$^{4}$ $\&$  A. Gomboc$^{1}$}
\affil{$^1$Faculty of Mathematics and Physics, University of Ljubljana, Jadranska ulica 19, SI-1000 Ljubljana, Slovenia \\
	     $^2$ Astrophysics Research Institute, Liverpool John Moores University, Liverpool, L3 5RF, UK\\
	     $^3$ Physics Departments, University of Ferrara, via Saragat 1, I-44122, Ferrara, Italy\\
	     $^4$ INAF – Osservatorio Astronomico di Brera, via E. Bianchi 46, 23807 Merate (LC), Italy}

\email{jure.japelj@fmf.uni-lj.si}

\begin{abstract}
We use a parent sample of 118 gamma-ray burst (GRB) afterglows, with known
redshift and host galaxy extinction, to separate afterglows with and
without signatures of dominant reverse-shock emission and to determine which physical conditions lead to a prominent reverse-shock emission. We identify 10 GRBs with reverse shock signatures - GRBs 990123, 021004, 021211, 060908,
061126, 080319B, 081007, 090102, 090424 and 130427A.
By modeling their optical afterglows with reverse and forward shock
analytic light curves and using Monte Carlo simulations, we estimate the
parameter space of the physical quantities describing the ejecta and
circumburst medium. We find that physical properties cover a wide parameter
space and do not seem to cluster around any preferential values.
Comparing the rest-frame optical, X-ray and high-energy
properties of the larger sample of non-RS-dominated GRBs, we show that the
early-time ($<$ 1ks) optical spectral luminosity, X-ray afterglow
luminosity and $\gamma$-ray energy output of our reverse-shock
dominated sample do not differ significantly from the general population
at early times. However, the GRBs with dominant reverse shock emission have fainter
than average optical forward-shock emission at late time ($>$ 10 ks). We find that GRBs with an identifiable reverse shock component show high magnetization parameter $R_{\mathrm{B}} = \varepsilon_{\rm B,r}/\varepsilon_{\rm B,f} \sim 2 - 10^4$. Our results are in agreement with the mildly magnetized baryonic jet model of GRBs.

\end{abstract}


\section{Introduction}
\label{s1}
The study of gamma-ray burst (GRB) afterglows started with their discovery in 1997 \citep[][]{costa1997,paradijs1997}. Since then, afterglow observations have established the cosmological nature of GRBs \citep[e.g.,][]{gomboc2012}, provided information on their stellar progenitors \citep{hjorth2012} and prompted the study of GRB circumburst environment \citep[e.g., ][]{petitjean2011}, their host galaxies \citep[][]{berger2011, fynbo2012}, and intergalactic medium in the GRB line-of-sight \citep[e.g., ][ and references therein]{vergani2009, chornock2013}.

Afterglow emission has also been considered as a powerful probe capable of revealing physical properties in gamma-ray burst ejecta as well as the medium through which the fireball propagates. According to the standard afterglow model \citep{sari1995,meszaros1997}, a relativistically expanding fireball propagating through a medium surrounding a GRB progenitor drives a shock into the medium, known as a forward shock (FS). Heated electrons behind the shock emit synchrotron radiation, giving rise to FS afterglow emission \citep{sari1998}. In addition to the forward shock, a reverse shock (RS), propagating back into the fireball, can be produced \citep{sari1999b}. 

The FS afterglow model has proven to describe late-time afterglow behavior well.  The environmental dependence of afterglow light curve evolution has been calculated analytically for two limiting cases: constant density interstellar medium (ISM) \citep[e.g., ][]{sari1998} and stellar wind environment with a density profile $\propto r^{-2}$ \citep[e.g., ][]{chevalier2000}, where $r$ is distance from the progenitor. The distinct temporal and spectral behavior of the two environments can be used to determine the nature of the medium surrounding the progenitors \citep{schulze2011} using predicted relations between  temporal and spectral afterglow slopes, i.e., closure relations \citep[e.g., ][]{zhang2006}. However,  in order to constrain the values of the microphysical parameters in the ejecta (the ratio of the electron and magnetic energy density over the internal energy density in the shocked region, $\varepsilon_{\mathrm{e}}$ and $\varepsilon_{\mathrm{B}}$, electron energy distribution index $p$), an analysis of the FS emission alone is insufficient. Emission from the RS can be used to measure the values of microphysical parameters,  to derive a Lorentz factor of the ejecta, and to constrain the nature of the ejecta itself \citep{zhang2003,kobayashi2003}. 

The evolution of the RS emission has been mostly studied for two limiting cases \citep{kobayashi2000}: $(i)$ the thick shell case, in which a RS becomes relativistic and starts to decelerate shell material, and $(ii)$ the thin shell case, in which a RS stays sub-relativistic and is too weak to decelerate the shell. Observational evidence suggest that the two limiting cases might not describe the real conditions very well \citep[e.g.,][]{virgili2013}. Intermediate events (between the relativistic and sub-relativistic cases) should be handled by numerical studies \citep{nakar2004,harrison2013}. All these studies assume a brief central engine activity and a constant Lorentz factor of the ejecta, resulting in a \textit{short-lived} reverse shock emission. However, ejecta could be composed of shells of different Lorentz factors. In such scenario, slower shells continue to feed the blast wave, giving rise to a \textit{long-lived} reverse shock emission. In this case, depending on the mycrophysics parameters, light curve can be completely dominated by a RS emission for the duration of the afterglow \citep[][]{uhm2007,genet2007}. This work focuses exclusively on the short-lived RS emission.

RS emission is expected to be especially prominent at low frequencies (optical to radio) and can be recognized by its characteristic rising and decaying slopes. For a typical set of micropysical parameters and initial Lorentz factor of the shell, the optical band lies between the typical ($\nu_{m,r}$) and cooling ($\nu_{c,r}$) synchrotron frequencies of the RS. In a constant ISM medium, the emission is predicted to reach its peak at the fireball deceleration time and then decay with a characteristic power-law slope of $\sim 2$ \citep{kobayashi2000} for both thin- and thick-shell cases. The rising index should be very steep ($\sim -5$) for a thin-shell or shallow ($\sim -1/2$) in the thick-shell approximation. In the case of a wind medium, a shallow rise and a steeper decay slope of $\sim 3$ are expected (for standard parameters) for both thick- and thin-shell scenarios \citep{kobayashi2003b, zou2005}. Depending on the relative strength and peak times of a reverse and forward shock afterglow component, three different light curve configurations are expected to be observed \citep[][]{zhang2003,gomboc2009}: 
\begin{itemize}
\item Type I: light curve with prominent reverse and forward shock afterglow peaks,
\item Type II: light curve with characteristic flattening due to bright RS afterglow outshining the FS emission,
\item Type III:  light curve with simultaneous FS and RS peaks, where the former outshines the latter.
\end{itemize}
RS emission arising from mildly magnetized outflows is predicted to be highly polarized \citep{lyutikov2003, granot2003, rossi2004}. Polarimetric measurements of early-time afterglow emission can thus provide a complementary method of recognizing or confirming a RS emission component \citep{mundell2007a,steele2009,mundell2013}.

Before 2005, only a few afterglows had been observed less than $\sim 1$ hour after the GRB trigger, with RS components being identified in three of them (GRB\,990123 - e.g., \citealt{sari1999a}, \citealt{kobayashi2000}, \citealt{soderberg2002}; GRB\,021004 - e.g. \citealt{kobayashi2003}, \citealt{kobayashi2003b}; GRB\,021211 - e.g., \citealt{fox2003}, \citealt{wei2003}).

With the launch of NASA's \textit{Swift} satellite \citep{gehrels2004} and the advent of purpose-built rapid-response autonomous robotic telescopes (such as the Liverpool Telescope and Faulkes telescopes [LT/FT; \citealt{steele2004}], Rapid Eye Mount [REM; \cite{zerbi2001}], Robotic Optical Transient Search Experiment [ROTSE; \citealt{akerlof2003}], etc.), the possibility of routinely observing the very early afterglow of large numbers of GRBs became a reality. RS optical flashes were expected to be ubiquitous in this new era of rapid follow-up. Surprisingly, the rate of detected RS components has been extremely low \citep{roming2006,melandri2008,gomboc2009,oates2009}. The lack of detections has been attributed either to strongly magnetized outflows, which can suppress the RS emission \citep{zhang2005}, the RS emission peaking at lower frequencies than the optical band (e.g. IR/mm) at early time \citep{mundell2007b,melandri2010}, or prompt optical emission originating in an internal shock region outshining  any contemporaneous external RS emission \citep{kopac2013}. 

To better understand the nature of RS emission, we compile a sample of GRBs with optical afterglows which show RS signatures. We compare their rest-frame optical, X-ray and $\gamma$-ray properties to a larger sample of GRBs with no apparent RS contribution in their optical afterglow. To investigate whether the physical conditions in the GRB ejecta of our RS afterglow sample show similar or different properties, we use a simple analytical model of reverse- and forward-shock emission and apply it to our RS sample using Monte Carlo simulations. We examine a parameter space that describes our afterglows well and discuss relations between various parameters.

Throughout the paper the convention $F_{\nu,\mathrm{t}} \propto t^{-\alpha}\nu^{-\beta}$ is adopted, where $\beta$ and $\alpha$ are spectral and temporal afterglow slopes, respectively. Standard cosmology with $H_{0} = 71~\mathrm{km}~\mathrm{s}^{-1}~\mathrm{Mpc}^{-1}$, $\Omega_{M} = 0.3$ and $\Omega_{\Lambda} = 0.7$ is assumed.  Times are given with respect to GRB trigger time.

\section{Sample Selection and Broad-band Data}
\label{s2}
In order to compare the rest-frame properties of GRBs with and without RS contribution in optical wavelenghts, we compile a comprehensive sample of GRBs with measured redshift. For a GRB to be included in the sample, it must have available optical/NIR afterglow observations with data published or submitted up to Sep 2013 in refereed journals (GRBs with data published only in GCNs\footnote{http://gcn.gsfc.nasa.gov/} are not included in our sample). Furthermore, in order to calculate rest-frame luminosities, knowledge of the optical spectral index ($\beta_{\mathrm{O}}$) and host galaxy extinction in the line-of-sight ($A_{\mathrm{V}}$) is required. A sample of 118 GRBs (27 detected in the pre-\textit{Swift} era) satisfy all of the above requirements. GRBs in this sample are summarized in Table \ref{tab1} in the Appendix.

\subsection{Optical data}
\label{s20}
Photometric measurements were collected from published data. For observations carried out with multiple filters showing no significant color evolution in the afterglow, we shift all light curves to a common band (using the measured spectral index $\beta_{\mathrm{O}}$) to achieve improved time coverage. Rest-frame extinction of some intermediate-redshift bursts is rather high. In those cases, we use NIR data, if available, in order to reduce uncertainty due to host extinction correction.  Where necessary, we correct late-time light curves for host contribution. We do not include data contaminated by supernova emission. We correct observed magnitudes for Galactic extinction assuming Galactic extinction maps provided by \citet{schlegel1998} and average extinction profile given in \citet{cardelli1989}. Magnitudes are converted to flux densities using proper filter-dependent zero-magnitude fluxes for calibration \citep{fukugita1995, fukugita1996}. Flux densities are further corrected for host extinction $A_{\mathrm{V}}$, using values and extinction laws reported in Table \ref{tab1}. Knowing the spectral slope of an afterglow, its monochromatic light curve, observed at frequency $\nu$ at time $t_{\rm obs}$ after a GRB trigger time, is  transformed to a rest-frame spectral luminosity using:
\begin{equation}
\label{eq1s1}
L_{\nu_{\mathrm{R}}}(t_{\mathrm{rest}}) = \frac{4\pi d_{\mathrm{L}}(z)^{2}}{\left( 1 + z\right)^{1 - \beta_{\mathrm{0}} + \alpha}}F_{\nu}(t_{\rm obs})\left( \frac{\nu_{\mathrm{R}}}{\nu}\right)^{-\beta_{\mathrm{O}}},
\end{equation}
where $d_{\mathrm{L}}(z)$ is the luminosity distance, $F_{\nu}$ is the flux density corrected for host and Galactic extinction, and $t_{\mathrm{rest}}$ is time measured in the GRB rest frame. Luminosities are calculated at the rest-frame frequency $\nu_{\mathrm{R}}$ corresponding to the Cousin $R$ filter. 

\subsection{X-ray data}

After the launch of the \textit{Swift} satellite, afterglow observations in the 0.1 - 10 keV energy range with the XRT telescope \citep{burrows2005b} became routine. Afterglow light curves observed with the XRT were studied extensively by \citet{margutti2013}, who report best-fit models to unabsorbed X-ray light curves in the rest-frame 0.3 - 30 keV energy range. We use their results to construct rest-frame X-ray light curves for 79 GRBs in our sample. These GRBs are flagged with a letter ``$B$" in the 9$^{\rm th}$ column of Table \ref{tab1}. The advantage of using these fitted light curves is that flares, commonly found in X-ray light curves \citep[e.g., ][]{burrows2005a,chincarini2007}, have been removed in the fitting procedure. 

\subsection{High-energy data}

We collected prompt $\gamma$-ray properties, namely the duration of the prompt burst ($T_{90}$) and isotropic equivalent emitted energy ($E_{\mathrm{\gamma,iso}}$), from the literature (values and references are reported in Table \ref{tab1}). Most of the energy values are reported for emitted energy in the rest-frame range of 1-10$^4$ keV. Values which have not been calculated for this particular energy range,are marked as lower limits (in all those cases the energy range is within 1-10$^4$ keV limits). $T_{90}$, corresponding to the time in which 90$\%$ of the burst fluence is recorded, is energy dependent (i.e., $\gamma$-ray emission observed in different energy bands lasts for different time periods; \citet{virgili2012,qin2013}). Since GRBs in our sample have been detected with different instruments with different spectral characteristics, the reported $T_{90}$ values are in general calculated for different energy bands. 

\subsection{Radio data}

We note that five of the RS candidates in our sample have a detected radio afterglow (GRB\,990123 - \citealt{kulkarni1999b}; GRB\,021004 - \citealt{kobayashi2003}; GRB\,080319B - \citealt{racusin2008}; GRB\,090424 - \citealt{chandra2009}; GRB\,130427A - \citealt{laskar2013}). Although we do not discuss detailed radio properties, we use the radio detections in Section 4 where we apply theoretical models to the observed RS sample afterglow light curves.

\begin{deluxetable}{lcccr}  
\tablecolumns{5}
\tablecaption{Sample properties}
\tablehead{   
  \colhead{Sample} &
  \colhead{N} &
  \colhead{Early detection} &
  \colhead{Reverse} &
  \colhead{Non-reverse}
}
\startdata
Sample A & 118 & 79 & 10 & 36\\ 
Sample B & 79  & 63 & 6 & 34
\enddata
\tablecomments{Summary of the two samples used in the paper: a full sample of 118 GRBs (Sample A) and a subsample of 79 GRBs which were observed with the \textit{Swift} XRT instrument and analyzed by \citet{margutti2013} (Sample B). For each sample we report the number of GRBs with early ($t_{\mathrm{rest}} < 500$ s) optical afterglow observations, the number of RS candidates, and the number of GRBs for which we do not find evidence for RS emission in optical afterglow.}
\label{tab0}
\end{deluxetable}

\subsection{Selection of GRBs with reverse shock contribution}
\label{sample}
We constructed two samples from our parent sample:
\begin{itemize}
\item Sample A with all 118 GRBs,
\item Sample B with 79 GRBs with both optical and \textit{Swift} XRT detection, whose XRT data were analyzed by \citet{margutti2013}.
\end{itemize}
The two samples are summarized in Table \ref{tab0}. Evidence for a possible RS signature must be looked for at an early stage of optical afterglow emission. In the third column in Table \ref{tab0} we report the number of afterglows where optical emission was detected earlier than 500 s after the start of the GRB in the rest frame. In the last column we report that 36 (34) GRBs in Sample A (B) show no evidence of RS emission despite a well sampled early-time light curve. As discussed in the Introduction, that does not necessarily imply a complete absence of a RS component. Early optical afterglow light curves of the remaining GRBs show complicated emission components, which cannot be easily classified in the context of purely forward- and reverse-shock emission. We consider an afterglow as a RS-sample candidate if its light curve resembles the Type I or Type II morphology. We base our final decision on single-burst studies, where a detailed analysis confirms, or at least does not disprove the existence of RS emission component.

In our full sample (A) we have 10 afterglows that show evidence of an optical RS contribution: GRB\,990123, 021004, 021211, 060908, 061126,  080319B, 081007, 090102, 090424 and 130427A. For five of them a RS component has been firmly confirmed (GRB\,990123 - e.g., \citealt{sari1999a}; GRB\,021211 - e.g., \citealt{fox2003}, \citealt{wei2003}; GRB\,061126 - \citealt{gomboc2008}; GRB\,081007 - \citealt{jin2013}; GRB\,130427A - \citealt{laskar2013,perley2013b}). However, the remaining five either lack  good early-time photometric coverage (GRB\,060908 - \citealt{covino2010}; GRB\,090102 - \citealt{gendre2010,steele2009}; GRB\,090424 - \citealt{jin2013}) or have different interpretations (GRB\,080319B -  RS \citep{bloom2009} vs. two-component jet model \citep{racusin2008}; GRB\,021004 - RS \citep{kobayashi2003,kobayashi2003b} vs. multiple energy injections \citep{postigo2005}). GRB 021004 is the only case in our sample with a possible Type I light curve, while the other nine are Type II. Light curves of the ten afterglows in Sample A are discussed in detail in Section \ref{s33}.

There are other cases of possible RS afterglows (GRB\,060111B - \citealt{klotz2006}, \citealt{stratta2009}; GRB\,060117 - \citealt{jelinek2006}; GRB\,091024 - \citealt{virgili2013}), which do not have measured redshift and/or optical spectral slopes and therefore have not been included in our sample. 

\begin{figure}[!]
\epsscale{1.25}
\plotone{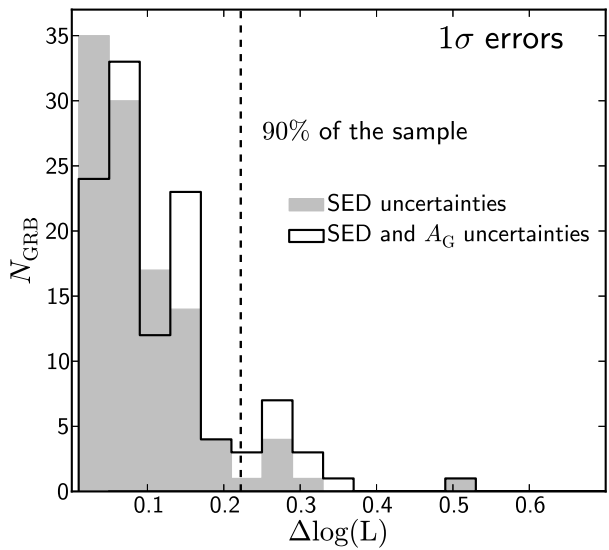}
\caption{Afterglow luminosity uncertainty estimates due to SED uncertainties (filled histogram) and with added Galactic extinction uncertainties (empty histogram). Most of the values of the final estimates are below 0.25 dex. }
\label{fig1}
\end{figure}

\subsection{Optical luminosity caveats and uncertainties}
\label{s21}
Optical luminosity light curves, obtained using the outlined procedure, are subject to a number of uncertainties due to a lack of precise knowledge of afterglow spectral behavior. The first contribution to the error comes from the measured quantities $A_{\mathrm{V}}$ and $\beta_{\mathrm{O}}$. The values we compiled in Table \ref{tab1} were obtained with a standard procedure, i.e., by fitting the afterglow spectral energy distribution (SED) - in either broadband NIR to X-ray or only NIR to UV wavelength range - with a featureless power-law or broken power-law model to describe the afterglow continuum emission \citep[e.g.,][]{sari1998}, which is then extinguished by scattering and absorption of light on dust and gas in the GRB line-of-sight (e.g., \citealt{kann2006,kann2010}, \citealt{schady2007,schady2010}, \citealt{greiner2011}, \citealt{zafar2011}, \citealt{covino2013}, \citealt{zaninoni2013}). When X-ray data are not included in analysis, a degeneracy between the spectral slope and the extinction is harder to break, which can result in less accurate parameter values \citep{covino2013}.

\begin{figure*}[!]
\epsscale{1}
\plotone{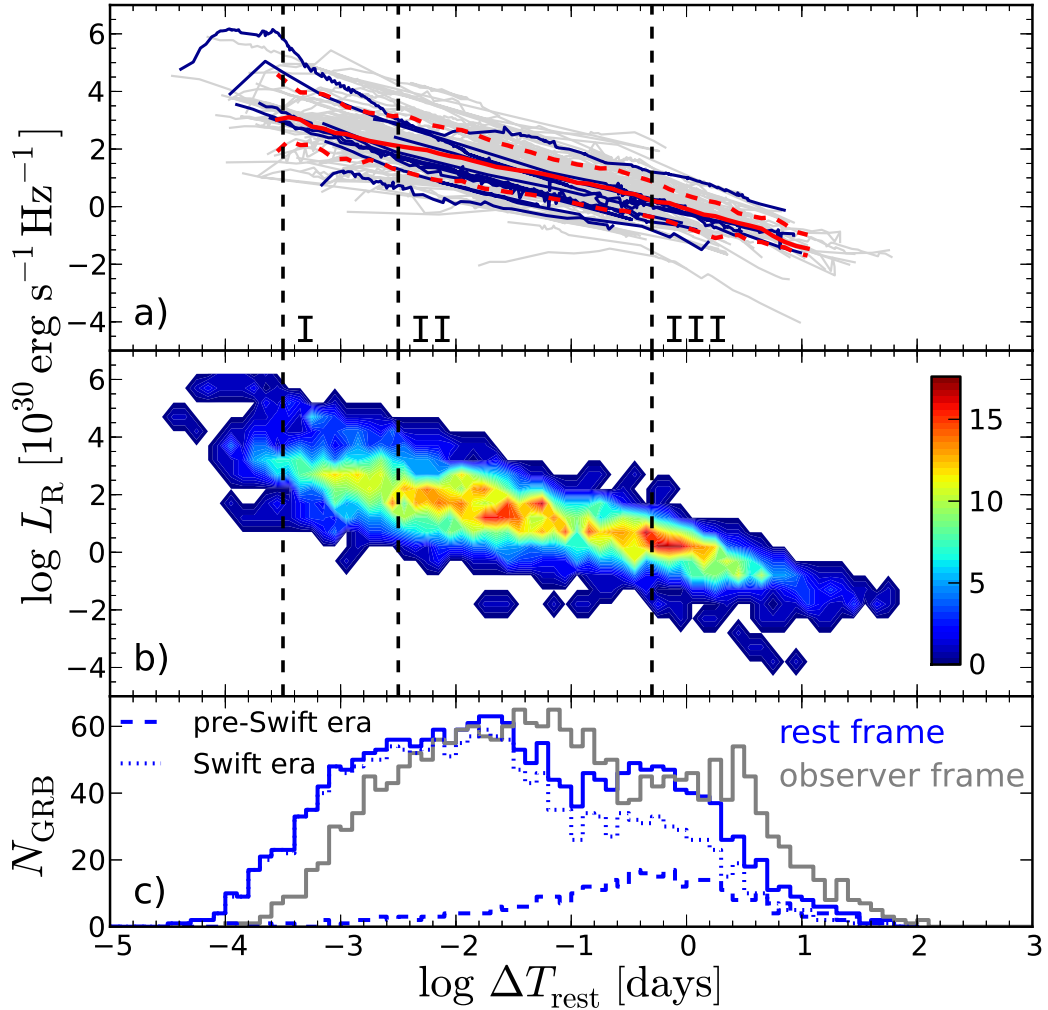}
\caption{$R$-band spectral luminosity light curves of our sample of 118 GRBs. $\Delta T_{\mathrm{rest}}$ is the time measured relative to the start of the GRB $\gamma$-ray emission and given in the GRB rest frame. Observed fluxes have been corrected for Galactic and host extinction and transformed to luminosities as described in Section \ref{s20}. (a) Full and dashed red lines represent the mean and 25$\%$ and 75$\%$ quartiles of the spectral luminosity distribution at each time bin. Spectral luminosity distributions at the three epochs, marked with dashed vertical lines, are plotted in Fig. \ref{fig3}. Afterglows with RS contribution are plotted in blue. (b) Contour plot showing the spectral luminosity distribution - the color scale represents the number of afterglows in a specific bin (0.1 dex in time and 0.5 dex in luminosity space). (c) Number of afterglows in a specific time bin (0.1 dex) in rest (blue) and observer frame (gray); the contribution of pre-\textit{Swift} (dashed) and \textit{Swift}-era (dotted) afterglows are also shown. We use the actual observed data to create this plot. However, in the subsequent analysis we use light curve models, fitted on the measurements (see Section \ref{s3}).}
\label{fig2}
\end{figure*}

We perform a Monte Carlo (MC) simulation to estimate the luminosity uncertainty due to errors in the measurements of $A_{\mathrm{V}}$ and $\beta_{\mathrm{O}}$. In order to make a realistic estimate, we take into account the degeneracy between the two parameters. We assume that both parameters are distributed according to a bivariate normal distribution\footnote{Since we do not know the actual distribution of the two parameters, we assume the normal distribution as the most natural choice.}, where the correlation coefficient $\rho$ is estimated using the SED fit results of \citet{kann2010}. We measure a Pearson correlation coefficient of $\rho = 0.35 \pm 0.15$ and adopt this value for our analysis, estimating the errors by performing a MC simulation in which we take into account the errors of the best-fit parameters. Once we randomly draw both $A_{\mathrm{V}}$ and $\beta_{\mathrm{O}}$, we calculate the luminosity from Eq.\@ (\ref{eq1s1}) and its difference $\Delta \log L$ from best-fit parameters value (logarithmic values are used throughout the paper). We repeat the outlined procedure for 10,000 times for each afterglow. Simulated differences $\Delta \log L$ are distributed according to a normal distribution centered at 0. By fitting the distribution we obtain a standard deviation of the distribution, which is our 1-$\sigma$ error luminosity estimate for a particular afterglow. A distribution of error estimates for Sample A is shown in Figure \ref{fig1} (filled histogram).

We consider another potential source of error: the correction for Galactic extinction. Galactic extinction maps provided by \citet{schlegel1998} are used to correct the data and play an important role in the derivation of $A_{\mathrm{V}}$ and $\beta_{\mathrm{O}}$ for almost all GRB afterglows analyzed in the literature. In order to be consistent, we use the same maps in our work. However, by analyzing the colors of stars observed with the Sloan Digital Sky Survey, \citet{schlafly2010} and \citet{schlafly2011} find that the values reported by \citet{schlegel1998} are systematically overestimated by approximately $14\%$. Another problem is the use of a total-to-selective extinction ratio $R_{\mathrm{V}} = 3.1$ for the conversion of reddening to absolute extinction, since this is only an average value of an otherwise rich ensemble of values corresponding to different lines-of-sight. For example, examining a few hundred lines-of-sight \citet{fitzpatrick2007} found $R_{\mathrm{V}} = 2.99 \pm 0.27$. We estimated $(i)$ the relative error one obtains in calibration by using overestimated Galactic extinction values and $(ii)$ the relative error due to the dispersion of $R_{\mathrm{V}}$ values by using a MC simulation. Combining both errors in quadrature reveals that the combined uncertainty in flux calibration does not rise over $\approx 10\%$ for most of the sample. However, this error contribution is more or less negligible when compared to the uncertainties in measuring $A_{\mathrm{V}}$ and $\beta_{\mathrm{O}}$. This is clearly shown in Figure \ref{fig1} where we plot the combination of both effects with a solid black line. Most of the values\footnote{Errors reach much higher values in a few cases mostly due to large uncertainties in derived extinction values (see Table \ref{tab1}).} are below 0.25 dex, which we take as a reference uncertainty estimate in this study.

While there are some afterglows with multiepoch SED analyses (e.g., 080319B), most afterglows in our sample had $\beta_{\mathrm{O}}$ measured only at one epoch, thus we cannot account for spectral evolution when calculating rest-frame luminosities (see Eq.\,\ref{eq1s1}). To understand the magnitude of this effect, we calculate how much the luminosity values would change if the real spectral slopes differed from those reported in Table \ref{tab1} by $\Delta \beta = 0.5$ \citep[e.g., as expected from the passing of the cooling frequency through the observing band;][] {sari1998}. For most cases in our sample, the corresponding change is $\approx 0.25$ dex.

Throughout the paper we use isotropic equivalent luminosities, as the beaming angles are unknown for most of the GRBs in the sample. The interpretation of late-time properties (Section \ref{s3}) and modeling of late-time data (Section \ref{s33}) could be wrong if the steepening of the light curves that is due to relativistic beaming effect \citep{rhoads1997,sari1999c}, is not taken properly into account. We discuss the implications in Sections \ref{s31} and \ref{details}.

\section{Results and discussion}
\label{s3}

Rest-frame $R$-band spectral luminosity light curves of afterglows from our sample, corrected for Galactic and host extinction, are plotted in Figure \ref{fig2}a. Figure \ref{fig2}b shows light curves binned with a temporal step of 0.1 dex and luminosity step of 0.5 dex - the latter is based on the error estimate discussed in Section \ref{s21}. In the following, all times are given as rest-frame values, unless stated otherwise.

\subsection{Model Fitting}

Different temporal light curve sampling makes the qualitative comparison of afterglows difficult. To overcome this, we fit each afterglow light curve with a model, which is taken to be a power-law, a multiple-broken power-law \citep[e.g.,][]{beuermann1999,granot2002} or a linear combination of the two. It is not our goal to test the overall properties of the sample objects within the context of our model, rather to present a detailed light curve study tailored to the fine details of each individual GRB. Therefore, we do not provide fit results for each case \citep[detailed sample studies have been performed by e.g.,][]{zeh2006,li2012,zaninoni2013}. We also investigate whether using fitted light curves or original data change sample properties (e.g., luminosity distribution, see Section \ref{s31}). We find that there are no significant changes and in the following we use fitted model afterglows.

Next, we examine the luminosity properties of the 27 pre-\textit{Swift} GRBs to see whether they differ from the population of \textit{Swift}-era GRBs. As can be seen from Figure \ref{fig2}c, pre-\textit{Swift} afterglows mainly populate the late-time part of the plot and are expected to be biased toward brighter afterglows. However, using a two-sample Kolmogorov-Smirnov test on the spectral luminosity distributions of the pre-\textit{Swift} and \textit{Swift} afterglows in the time interval of $-1 \leq \log \Delta T [\mathrm{days}] \leq 1$ we obtain probabilities higher than $P = 0.7$ that the two populations are drawn from the same distribution. A similar result is obtained using a two-sample Anderson-Darling (AD) test ($P \sim 0.6$ in that time interval). Those two tests suggest that the populations are drawn from the same distribution. Additionally, the mean luminosity as a function of time (see Section \ref{s31}) is practically identical for \textit{Swift} and pre-\textit{Swift} populations in this time interval. With this in mind, we treat the two populations as one in the subsequent analysis. 

\subsection{Optical spectral luminosity distribution - the general case}
\label{s31}
We first investigate the time-dependent spectral luminosity distribution of all Sample A afterglows. For each time bin we compute the mean and two quartiles (25$\%$ and 75$\%$) of the distribution, which are plotted in Figure \ref{fig2}a with solid and dashed red lines, respectively. The time dependency of the mean itself can be represented with a broken power-law \citep{beuermann1999}. By fitting the data and assuming a smoothness parameter of $n = 1$ we get: $\alpha_{1} = 0.81 \pm 0.02, \alpha_{2} = 1.71 \pm 0.12$ and $t_{\mathrm{b}} = 2.34 \pm 0.35$ days (with $\chi^2/\mathrm{dof} = 36/42$). To check whether the late-time steepening, which could in principle be attributed to a jet break\footnote{Only a handful of GRBs have observed achromatic jet breaks \citep{liang2008} as predicted by a standard theory, therefore the identification and confirmation of a jet break is not trivial.}, has an impact on derived properties for the general population, we look for late-time breaks. In the cases for which the steepening is found, we extrapolate pre-break light curves to late times assuming a constant decay index. Afterglow light curves with only late-time observations are discarded. We find no statistical difference between jet-corrected sample and the original sample for times $\Delta T <$ a few days.

\begin{figure}[!]
\epsscale{1.25}
\plotone{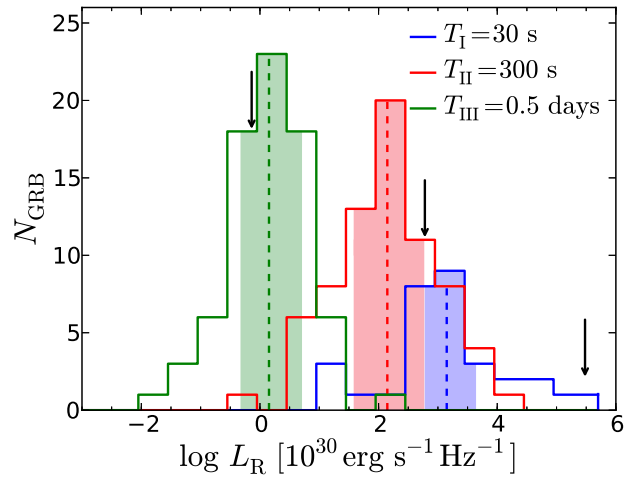}
\caption{Spectral luminosity distributions at three different epochs. Vertical dashed lines indicate the mean values of the distributions. Shadowed regions show the area where 50$\%$ of the data within each distribution lies (e.g., the area between 25$\%$ and 75$\%$ quartiles). The spectral luminosity of GRB\,080319B at the three epochs is marked with arrows.}
\label{fig3}
\end{figure}

\begin{figure}[!]
\epsscale{1.3}
\plotone{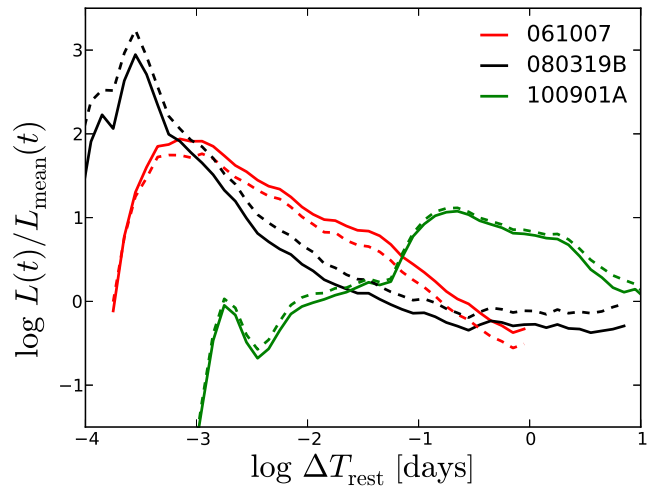}
\caption{Luminosity light curves of three GRB afterglows, \textbf{divided by} the mean luminosity light curve (see red solid line in Figure \ref{fig2}a). Dashed lines show the light curves which are not corrected for host extinction. The latter are divided by the mean light curve of the extinction-uncorrected data. Uncertainty, estimated in Section \ref{s21}, is $\sim 0.25$ dex.}
\label{fig4}
\end{figure}

Luminosity distributions at three epochs (as marked in Figure \ref{fig2} with dashed vertical lines) are shown in Figure \ref{fig3}. The rest-frame luminosity distribution has previously been investigated, with some works finding evidence of a bimodal distribution at late times \citep{nardini2006,nardini2008, liang2006, kann2006}. Later studies on smaller, more homogeneous samples \citep{melandri2008,oates2009, cenko2009}, as well as large sample studies (\cite{kann2010,zaninoni2013}), do not find significant bimodality. As suggested by Figure \ref{fig3}, we also find no evidence for late-time bimodality.

\begin{figure*}[!]
\epsscale{1.15}
\plottwo{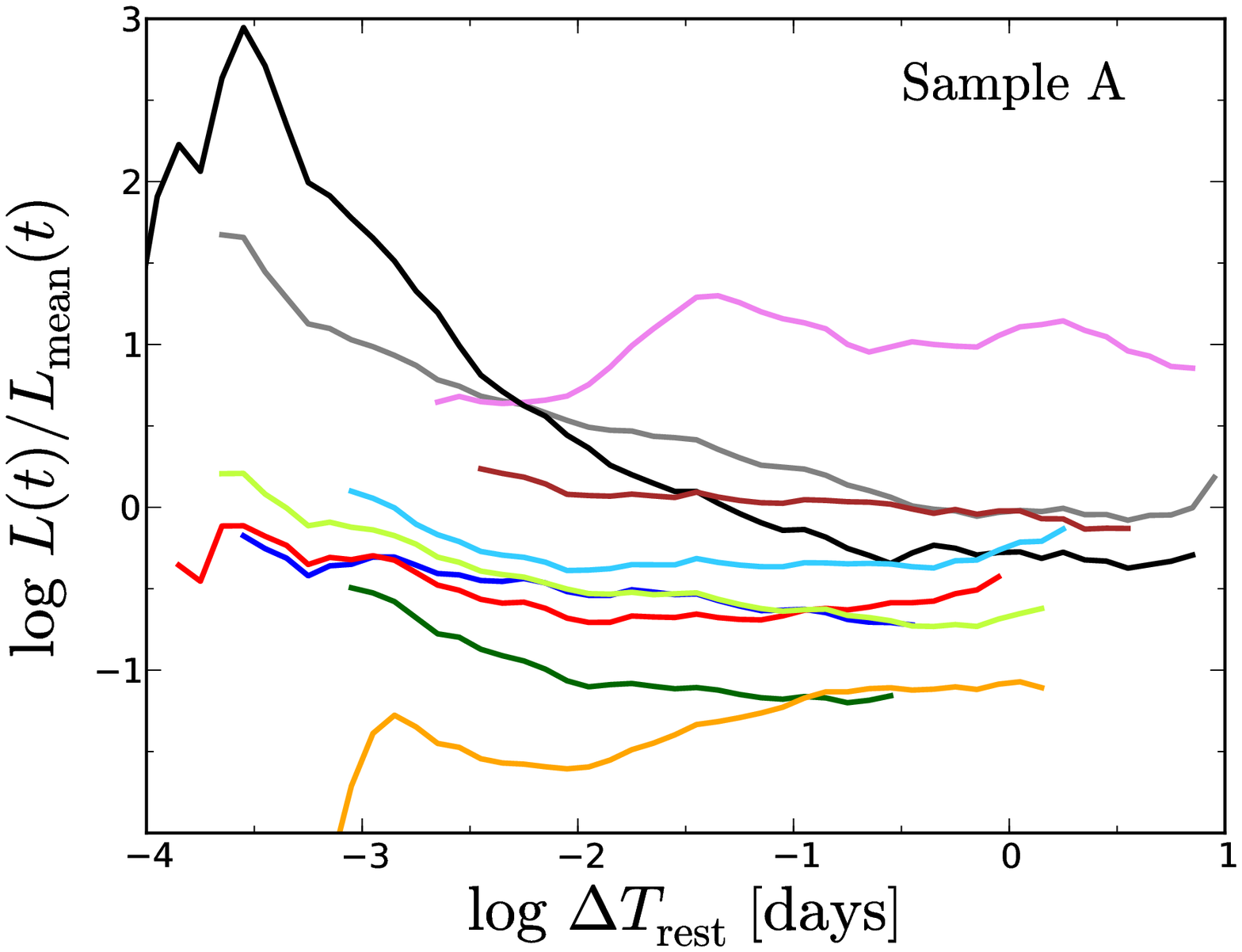}{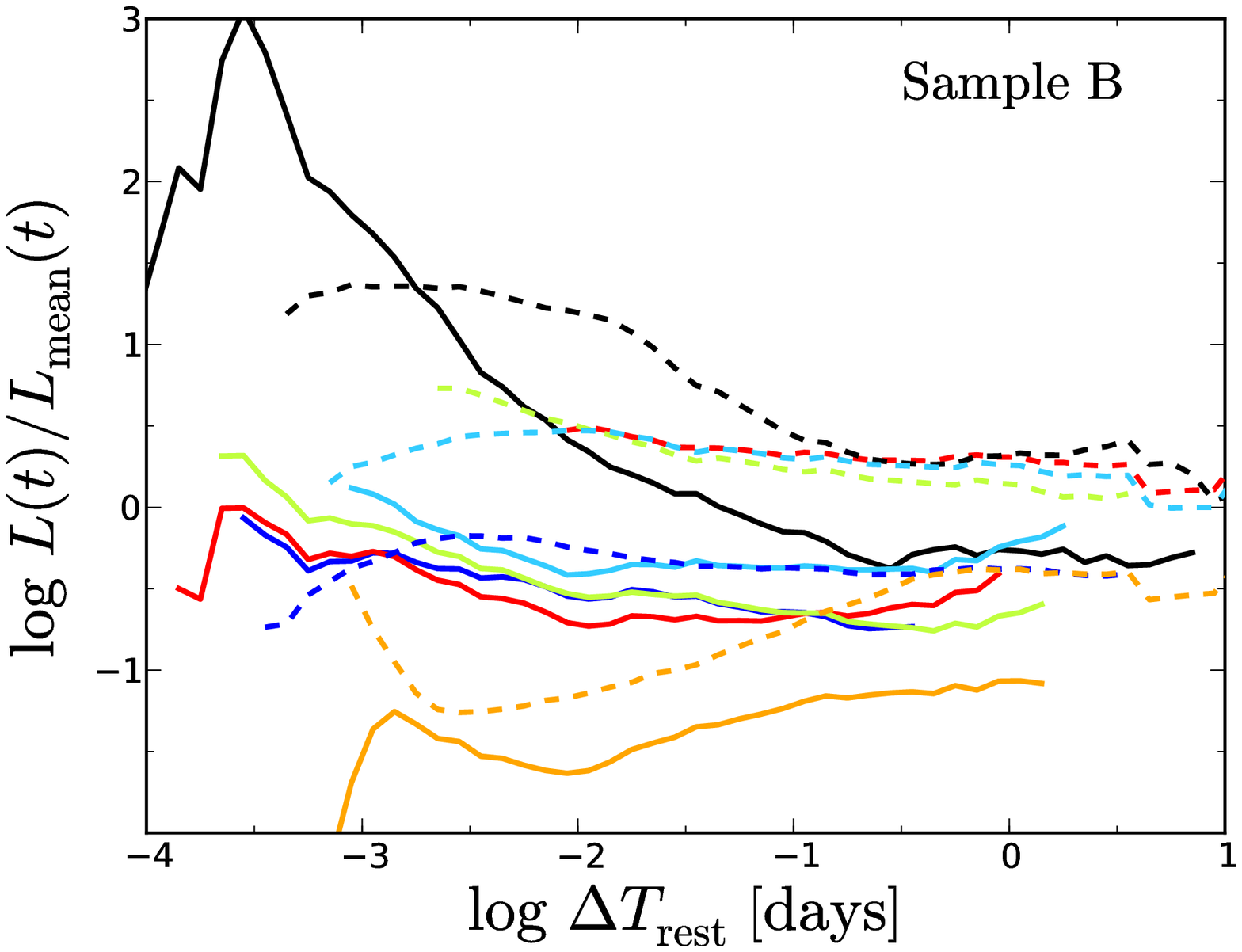}
\caption{Luminosity light curves of RS candidates, \textbf{divided by} the mean light curve of the sample, calculated for Sample A (left) and Sample B (right). Coloring for both figures: GRB\,990123 - gray, 021004 - violet, 021211 - dark green, 060908 - dark blue, 061126 - red, 080319B - black,  081007 - orange, 090102 - light green, 090424 - light blue, 130427A - brown. In the case of Sample B, normalized X-ray band light curves have been also calculated and are plotted with dashed lines. Uncertainty, estimated in Section \ref{s21}, is $\sim 0.25$ dex.}
\label{fig5}
\end{figure*}

Luminosities at early times are more dispersed than at late times. This can be seen both from the calculated standard deviation as well as the interval between the 25$\%$ and 75$\%$ quartiles of the distributions: both quantities are decreasing with time. A number of effects could be the cause for the larger early-time dispersion: different emission components (e.g., forward- or reverse-shock afterglow emission, internal-shock emission) or unaccounted spectral evolution when calculating the luminosities (see Eq. 1 and discussion in Section \ref{s21}).

Comparing distributions at different epochs is not a trivial task, since not all afterglows cover all time intervals. Another more subtle problem is that afterglow light curves show a wide variety of properties, with various decay slopes \citep[e.g.][]{oates2009} and features that can change their temporal evolution, like late-time rebrightenings \cite[e.g.][]{monfardini2006,nardini2011,greiner2013,gomboc2013} or density bumps \citep[e.g.][]{lazzati2002, guidorzi2005}. Consequently, the relative position of afterglows in the luminosity distribution changes with time. An example is shown in Figure \ref{fig3}, where the arrows point to the value of spectral luminosity of GRB\,080319B at the three chosen epochs. In the following we therefore plot afterglow light curves which are divided by the ``mean light curve'' (solid red line in Figure \ref{fig2}a). This approach allows us to immediately evaluate the relative flux of an afterglow (e.g., brighter or fainter with respect to the mean) and its temporal behavior. We show this for three distinct cases in Figure \ref{fig4}. The \textit{naked-eye burst} GRB\,080319B \citep[e.g.][]{racusin2008,bloom2009}, the brightest afterglow ever observed at an early stage, initially decays very fast and behaves like an average afterglow at late time. GRB\,100901A \citep{gomboc2013} is among the faintest at the beginning but experiences an extreme late-time rebrightening, making it among the brightest in the $\sim 0.1 - 1$ days time range. GRB\,061007 \citep{mundell2007b, rykoff2009} shows an early peak and a remarkably smooth decay without breaks, bumps, etc. However, its temporal decay index (steeper than the decay of the mean) and its absolute luminosity result in different time evolution with respect to the mean. 

\subsubsection{The Role of Host Extinction}
To show that the correction for host extinction is an important step in the analysis of rest-frame optical properties, we repeat the procedure, this time taking the data not corrected for host extinction. Results are shown as dashed lines in Figure \ref{fig4}. For 62$\%$ of the sample, the measured host extinction is low enough that the difference between the host extinction corrected and uncorrected light curves is within our error estimate (i.e., less than 0.25 dex, see Section \ref{s21}). For 19$\%$, the difference is larger than 0.5 dex. Measured rest-frame extinction values, $A_{\mathrm{V}}$, are generally not very high for most of the afterglows \citep[e.g.,][]{zafar2011}. However, due to relatively high redshifts, the light we observe in the optical band was actually emitted in the UV part of the spectrum, where light is considerably attenuated even at low dust quantities in the line-of-sight. The three most attenuated afterglows in our sample are GRB\,060210 \citep{curran2007,stanek2007}, GRB\,080607 \citep{perley2011} and GRB\,100621 \citep{kruhler2011}.

\subsection{Afterglows with and without RS emission}
\label{s32}
\subsubsection{Optical properties}
Normalized luminosity light curves of the Sample A RS candidates are shown in Figure \ref{fig5} (left). The afterglows are found to span five orders of magnitude in spectral luminosity at early times. The two bright afterglows decay rather fast, compared to the rest of the sample: after $\sim$ 1 day they behave like an average afterglow. This is in agreement with the result presented by \citet{oates2009,oates2012}, who found that the brighter the afterglow the faster it decays. The faint group, however, stays at the faint end of the distribution for most of the afterglow duration. The case of GRB\,021004 is curious - it is among the brightest afterglows at times $>1$ day, in complete contrast to the other RS afterglows.

First we check whether our RS sample is drawn from the same population as afterglows without a RS component. We compare our 10 RS candidates to the 36 Sample A afterglows with early-time observations and no compelling evidence of a RS using the two-sample KS test. The statistics are applied to each time bin, i.e., we compare the distribution of the two groups as a function of time. We find a KS probability $P_{\rm KS} \approx 0.65$ and $P_{\rm KS} \approx 0.06$ when comparing distributions at early time ($\log \Delta T [\mathrm{days}] < -2$) and late time ($\log \Delta T  [\mathrm{days}] > -1$), respectively. In addition, we estimate the error on the statistics by taking into account the estimated luminosity error and performing a MC simulation. The uncertainty does not reach over $\Delta P_{\rm KS} = 0.04$. We can thus reject the null hypothesis that the samples are the same at late times to the 90$\%$ confidence level. At early times, we cannot prove or disprove the hypothesis. We confirm this conclusion with the AD test. Due to the scatter in the brightness distribution at early times (see Figures \ref{fig2} and \ref{fig3}) it is not surprising that our RS sample is not significantly different from the rest of the population.  The late-time result, however, is more curious. GRB\,021004 seems to be an outlier. Its brightness at late time could be a result of multiple energy injections \citep{postigo2005}, a feature not recognized in any other RS-candidate light curve. We repeat the statistical analysis on the two samples excluding GRB\,021004. We obtain late-time probabilities $P_{\rm KS} < 0.03$ and $P_{\rm AD} < 0.04$ (including the uncertainty). Given these two results we can reject the hypothesis at the confidence level of 95$\%$. Apart from GRB\,021004 all afterglows with a reverse component are quite faint. Since the reverse component dominates only early-time emission, the result of the two tests suggests the possibility that the FS afterglows accompanied with reverse component are generally fainter. This could in principle be attributed to a selection effect: RS emission in the presence of a very bright FS could be masked by the FS emission.

\subsubsection{X-ray and $\gamma$-ray properties}

Observed preference toward fainter optical FS components might also reveal itself at higher energies. Six of our RS sample candidates have available X-ray light curves. None of the light curves shows strong evidence of a plateau phase: plateau, which is found in a large fraction of X-ray light curves \citep{nousek2006,zhang2006}, is a natural prediction of a long-lived RS model \citep[][]{uhm2007,genet2007}. A possible end of a plateau phase is observed in GRBs 060908, 090102, 090424 and 081007. GRBs 080319B and 130427A (the latter is not in the \citet{margutti2013} sample and therefore not included in ours) have no plateau phase, while the X-ray observations of GRB 061126 started too late to confidently exclude the presence of a plateau. We repeat the analysis we did on Sample A with Sample B, where in addition to optical, we also have X-ray light curves. The results are presented in Figure \ref{fig5} (right). We note that the mean light curve of the Sample B optical afterglows is very similar that of the Sample A afterglows, so the normalized optical light curves of the RS candidates are more or less unchanged. Normalized X-ray light curves are plotted as dashed lines. Late-time X-ray afterglows seem to be clustered in two groups. GRBs 061126, 080319B, 090102 and 090424 are among the brighter, while GRBs 060908 and 081007 are among the fainter group. However, they are all very near the mean X-ray light curve, given the much larger spread in late-time X-ray luminosities. 

\begin{figure}[!]
\epsscale{1.25}
\plotone{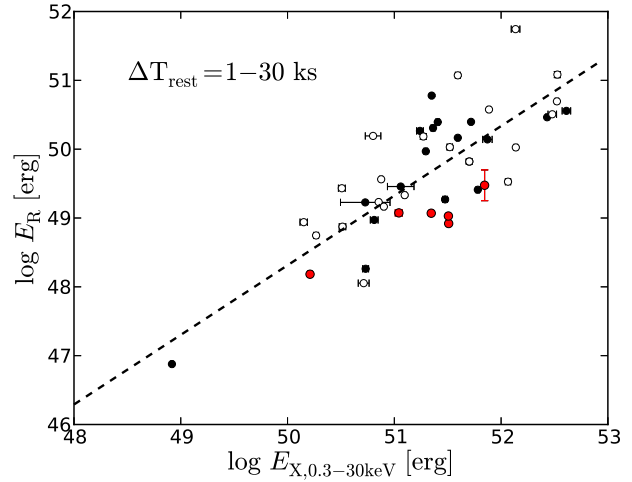}
\caption{Energy emitted in the $R$ band and 0.3-30 keV X-ray band in a rest-frame time interval of 1 - 30 ks. A total of 45 afterglows from Sample B were observed in that time interval. RS afterglows are plotted with red points. Afterglows with no apparent RS component are plotted with unfilled circles and the rest of the sample with black points. The line represents a power-law  fit to all measurements with a power-law index of 1.01 $\pm$ 0.12.}
\label{fig6}
\end{figure}

\begin{figure}[!]
\epsscale{1.25}
\plotone{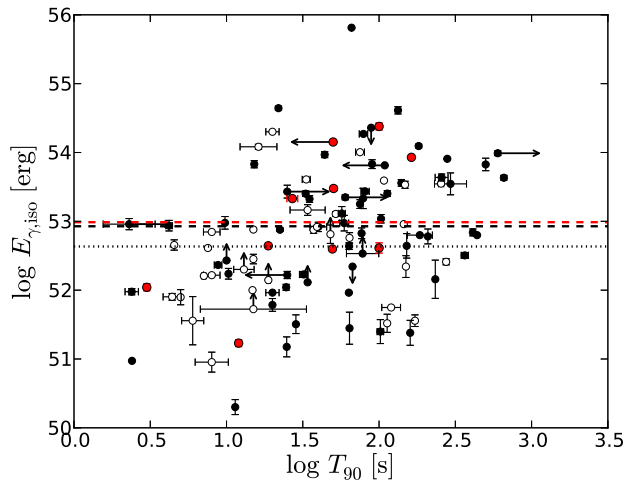}
\caption{High energy properties of our Sample A. Only GRBs with both $E_{\gamma,\mathrm{iso}}$ and $T_{90}$ measured are included. RS candidates are plotted with red points. Afterglows with no apparent RS component are plotted with unfilled circles and the rest of the sample with black points.  Upper/lower limits are indicated with arrows. Red and black dashed lines represent the median energy of the RS and overall sample, respectively. The black dotted line represent the median energy of afterglows without RS.}
\label{fig7}
\end{figure}

The above result can be investigated from the perspective of the Sample B afterglow energetics. To accomplish this, we first transform thr optical spectral luminosities to luminosities (multiplying by the $R_{\mathrm{c}}$ frequency window). We then choose a common time interval for the six RS candidates and as many other Sample B afterglows, attempting to maximize both optical and X-ray light coverage. The best compromise is to take the rest-frame time interval of $1 - 30$ ks in which a total of 45 afterglows have observations. We compute the energy emitted in that interval and plot it in Figure \ref{fig6}. In general we observe a clear correlation between the energy emitted in the two energy bands. We fit the data with a simple power-law function and obtain a slope of 1.01 $\pm$ 0.12. Afterglows with RS components are plotted in red and afterglows without a clear RS component are plotted with unfilled points. The latter have no preferential position in the plot. The six RS sample points are all below the best-fit power-law line. This could be a hint that X-ray afterglows accompanying RS candidates are relatively bright, in contrast to their optical counterparts. Unfortunately, only six RS afterglows in our sample fall in Sample B, preventing us from drawing any strong conclusion. In addition, we note that the X-ray light curves used in this analysis are integrated in the rest frame energy range of 0.3-30 keV, while optical light curves are obtained with observations through a relatively narrow frequency window.

Finally, we investigate $E_{\mathrm{iso}}$ and $T_{90}$, which are plotted in Figure \ref{fig7}. Equivalent isotropic energies of the RS GRBs do not show any preferential values. This is further confirmed using the KS and AD tests to compare the samples of RS and non-RS GRBs, where we obtain probabilities $P_{\rm KS} = 0.44$ and $P_{\rm AD} = 0.31$. We therefore cannot reject or confirm the hypothesis that the two samples come from the same distribution. We caution that the $T_{90}$ values are calculated for different energy ranges. For this reason we plot the values in Figure \ref{fig7} but do not use them for further analysis.

\section{Reverse Shock modeling}
\label{s33}

As discussed in the Introduction, reverse- and forward-shock emission components in an afterglow can be used to constrain the physical conditions in the GRB ejecta. Ideally, one would like to observe the behavior of both forward- and a reverse-shock emission, clearly distinguishing their respective afterglow peaks (i.e., Type I light curve). In this case, under the assumption that the electron distribution index, $p$, and the ratio of the electron energy density and the internal energy density in the shocked region, $\varepsilon_{\mathrm{e}}$, are the same in front of and behind the contact discontinuity, one can constrain the initial Lorentz factor, $\Gamma_{0}$, and the magnetization parameter $R_{\mathrm{B}}=\varepsilon_{\mathrm{B,r}}/\varepsilon_{\mathrm{B,f}}$, where $\varepsilon_{\mathrm{B,r}}$ and $\varepsilon_{\mathrm{B,f}}$ are the ratio of the magnetic energy density and the internal energy density in the reverse and forward shock region, respectively \citep{zhang2003}\footnote{Note that \citet{zhang2003} and \citet{zhang2005} define the ratio of magnetic field strength in the reverse and forward shock region as $R_{\rm B} = B_{\rm r}/B_{\rm f} = \left(\varepsilon_{\mathrm{B,r}}/\varepsilon_{\mathrm{B,f}}\right)^{1/2}$ while we assume $R_{\rm B} = \varepsilon_{\mathrm{B,r}}/\varepsilon_{\mathrm{B,f}}$ \citep[e.g.,][]{harrison2013}}. However, in most cases only one or neither peak is observed. Our RS sample mostly contains Type II light curves, where the RS emission dominates. \citet{gomboc2008}, extending the analysis of \citet{zhang2003}, showed how an approximate value of $R_{\mathrm{B}}$ in the case of Type II light curves can be estimated in the limiting case of thin- or thick-shell RS description. To estimate $R_{\mathrm{B}}$, the values of $\Gamma_{0}$, the shell deceleration time, and FS peak time have to be known. \citet{harrison2013} extended this analysis to the intermediate RS case.


Traditionally,  information on the physical properties of the GRB ejecta is inferred by fitting an empirical function to the observed multiwavelength light curve. Instead, we use an alternative approach in which we perform a Monte Carlo simulation to find a parameter space of light curves that best match the observed light curves, i.e., we assume the light curve is a combination of RS and FS emission and determine a set of physical parameters to reproduce it.

\subsection{A Simple Reverse plus Forward Shock Model}
We constructed a simple model of a reverse and forward shock afterglow. The connection of long GRBs with massive stars implies a circumburst environment in the form of a stellar wind. However, light curve and SED analysis of afterglows reveals that a constant-density ISM is a better approximation in majority of cases \citep[e.g.,][]{schulze2011} and the real conditions may be even more complicated (see discussion in Section \ref{radio}). Therefore, we decided to assume a constant ISM environment in our modeling, having in mind that this is only a rough approximation of real conditions (see also Section \ref{caveats}). Furthermore, we assumed a slow-cooling spectrum in which the typical synchrotron frequency, $\nu_{\mathrm{m}}$, is below the cooling synchrotron frequency, $\nu_{\mathrm{c}}$. In this case the spectrum is composed of three power-law segments: $F_{\nu} \propto \nu^{1/3}, \nu^{-(p-1)/2},  \nu^{-p/2}$, joined at break frequencies $\nu_{\mathrm{m}}$ and $\nu_{\mathrm{c}}$ \citep{sari1998}. Since we are primarily interested in optical wavelengths, we ignore synchrotron self-absorption (we look into this more carefully in Section \ref{radio}). Dependencies of  $\nu_{\mathrm{m}}$,  $\nu_{\mathrm{c}}$, and the peak flux in the spectral domain $F_{\nu,\mathrm{max}}$ (not to be confused with light curve peak $F_{\mathrm{p}}$) on the physical parameters and their temporal scalings were computed following equations in \citet{shao2005}, which are based on theoretical grounds described by \citet{sari1998} and \citet{kobayashi2000} for forward and reverse shock afterglows, respectively. For simplicity, we assume that only the thin-shell scenario applies to our data. Even though GRBs 990123 and 061126 are marginal cases (i.e., mildly relativistic),  they cannot heat the shell well and behave similarly to the thin-shell case \citep[e.g., ][]{kobayashi2000, gomboc2008}. 

The model light curve of an event occurring at redshift $z$ and observed at frequency $\nu$ is determined by a set of parameters 
\begin{equation}
F_{t,\nu,\mathrm{obs}} = f(t,\nu; z, p, n, \varepsilon_{B,f}, \varepsilon_{B,r},\varepsilon_{e}, E_{\mathrm{K}}, \Gamma_{0}),
\end{equation}
where $n$ is the density of circumburst ISM, $E_{\mathrm{K}}$ is the isotropic equivalent kinetic energy of the shell, and $\Gamma_{0}$ the initial Lorentz factor of the shell. Apart from $\varepsilon_{B}$ we assume the same values of microphysical parameters in the forward and reverse shock region. Our goal is to compare a theoretical model to the observed flux density optical light curves of our afterglows. Light curves are computed for $\nu = \nu_{\mathrm{R}}$. Redshift, $z$, is known for all afterglows in the sample. We are left with seven free parameters that have to be constrained by the actual observed light curve. We constrain the electron energy distribution index, $p$, by measuring the late-time (i.e., time when the contribution of the reverse component is negligible) FS afterglow decay index, $\alpha_{\mathrm{f}}$, and assuming the relation $\alpha_{\mathrm{f}} = 3(p_{\mathrm{f}} - 1)/4$ \citep{sari1998}, where we use the subscript $f$ to emphasize that $p_{\rm f}$ is measured from the FS decay slope. Since the RS decay is also dependent on $p$, we constrain $p$ to a very narrow interval around $p_{\mathrm{f}}$. This allows us to improve the modeling of RS decay slope in case the value of $p_{\mathrm{f}}$ does not provide a very good result.  We assume the parameters can take the following values: $p \in [p_{\mathrm{f}}-0.05,p_{\mathrm{f}}+0.05 ]$, $\varepsilon_{\mathrm{B,f}} \in [10^{-5}, 10^{-1}]$, $\varepsilon_{\mathrm{e}} \in [10^{-4}, 0.5]$, $n \in [10^{-1}, 10^{4}]$ cm$^{-3}$, $E_{\mathrm{K}} \in [10^{50}, 10^{56}]$ erg, $\Gamma_{0} \in [50, 10^4]$ and $\varepsilon_{\mathrm{B,r}} = R_{\rm B}\varepsilon_{\mathrm{B,f}}$, where $R_{\rm B} \in [1, 10^{5}]$. The latter assumption can result in an unphysical scenario with $\varepsilon_{\mathrm{B,r}} + \varepsilon_{\mathrm{e}} > 1$: such events are not considered in further simulation. We also assume that all parameters, apart from $p$, are uniformly distributed in log space. This is an arbitrary choice due to the lack of knowledge on the actual parameter distributions. As an additional constraint we place a limit on the radiative efficiency of the prompt gamma emission to be $\eta_{\gamma} > 0.01$ \citep{zhang2007}, thus placing the upper limit on the allowed $E_{\mathrm{K}}$ for each specific event. 

\begin{figure*}[!]
\epsscale{1.25}
\plotone{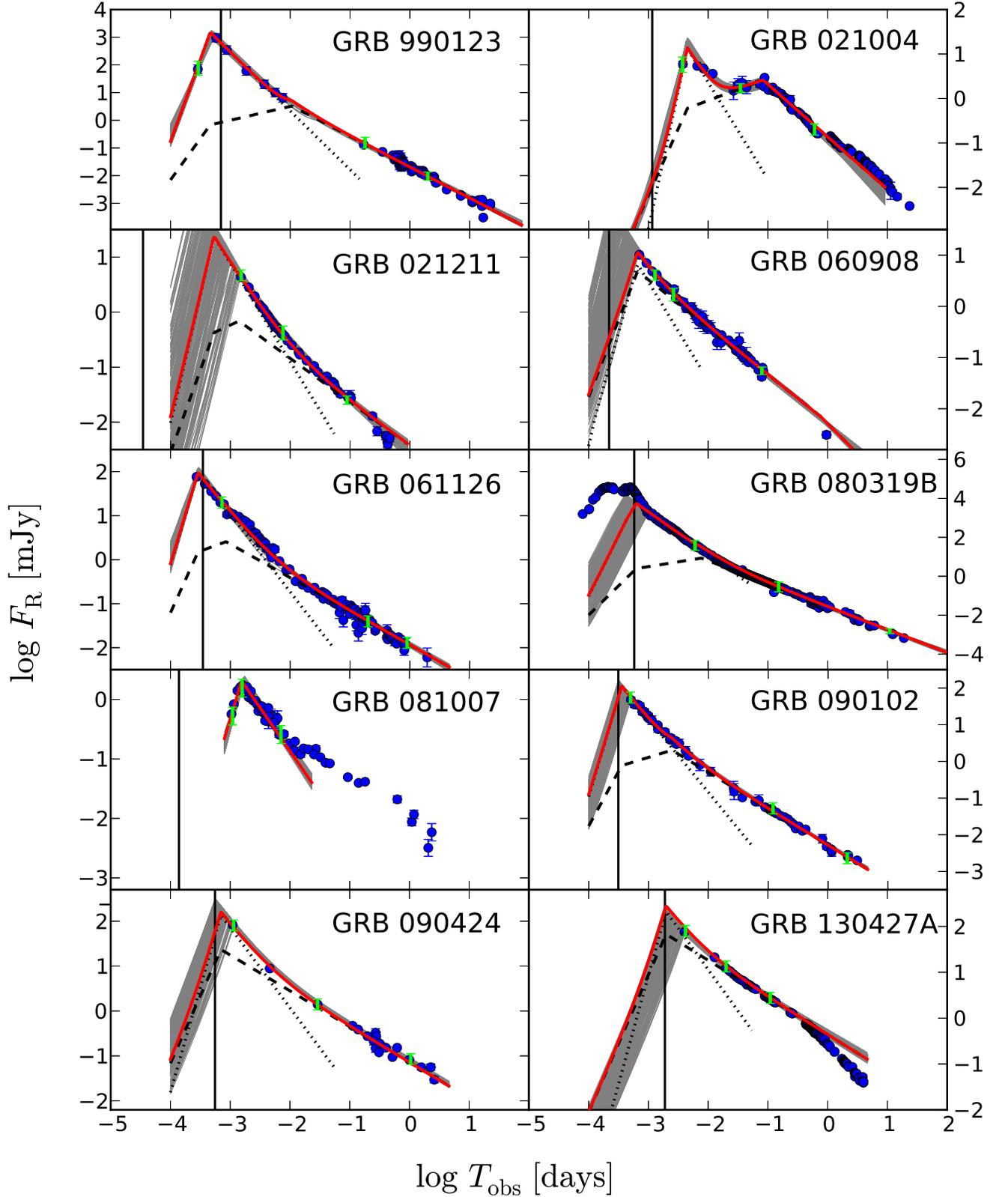}
\caption{Dust-unextinguished flux density light curves of our sample of 10 RS candidates in the observer frame. Observational data are plotted with blue points. Three green regions, in combination with rules reported in the text, have been used in the MC simulation to obtain theoretical models matching the data. 200 models for each GRB are plotted in gray and the best model among them is plotted in red. Each model is a combination of RS (black dotted lines) and FS (black dashed lines) emission. Vertical lines mark $T_{90}$ times as reported in Table 2. FS emission prior to the fireball deceleration time (i.e., RS peak time) is assumed to rise as $t^{3}$.}
\label{fig8}
\end{figure*}

By using monochromatic light curves, we expect the parameter space for each specific case will be only partially constrained. In addition, the assumption that the observed light curves can all be described as a combination of forward and reverse shock emission is a simplification which may not be entirely true (see later discussion for the need of additional emission components). For this reason, we do not attempt to find the best matching model (e.g., using $\chi^{2}$ statistics). Instead, we only need to find a sample of parameter sets, which can reproduce the observed light curves. By randomly choosing parameter values, we search for these models by imposing a few criteria the produced light curves have to meet: 
\begin{itemize}
\item If  a RS (FS) peak is observed, its peak flux $F_{\mathrm{p,r}}$ ($F_{\mathrm{p,f}}$) as well as its peak time $t_{\mathrm{p,r}}$ ($t_{\mathrm{p,f}}$) has to be reproduced within arbitrarily set accuracy interval ($\Delta \log F_{\mathrm{p}} < 0.2$ and $\Delta \log t_{\mathrm{p}} < 0.2$). In the thin-shell scenario and $\nu_{\mathrm{obs}} = \nu_{\mathrm{R}}$ we expect $t_{\mathrm{p,r}} \gtrsim T_{90}$. In practice, the last constraint has to be relaxed in cases when the peak is not observed and the first optical observation coincides with $T_{90}$: we assume $|\log t_{\mathrm{p,r}} -\log T_{90}| < 0.2$
\item Flattening, if observed, is characterized by its flux $F_{\mathrm{flat}}$ and time $t_{\mathrm{flat}}$ - in this case we require $ T_{90} < t_{\mathrm{p,f}} < t_{\mathrm{flat}}$. 
\item The last condition is normalization. We arbitrarily choose a few points on the observed light curves (it turns out that specifying flux density values at three different epochs is enough to obtain models, which can reproduce our light curves well), and allow a discrepancy in flux density values of 0.25 dex  between the data and the model.
\end{itemize}
With this procedure we search for the first 200 model light curves for each afterglow in our sample that satisfy the above requirements. This number of events was chosen to extract sufficient data for the analysis while maintaining a reasonable simulation execution time. This number does not affect the final conclusions. The best model among this 200 is then searched for using $\chi^2$ statistics. Observed host extinction corrected light curves and the corresponding best theoretical models are plotted in Figure \ref{fig8}. Parameter values of the best model for each case are reported in Table \ref{tab3}. 

\subsection{Modeling details}
\label{details}
For the case of GRB\,021004, we assume the rebrightening at $\sim0.1$ days is a FS peak. Early optical observations show a flattening instead of an expected RS peak - we assume the peak should have occurred somewhere between the first and the second observation.  We model only data with observations taken before 1 day, assuming that later rebrightening is due to another physical process and would thus need an additional component.

The last data point of the afterglow of GRB\,060908 (at $\sim 1$ day) is not included in the model: most possible models thus overpredict the flux density at late times, suggesting a jet break. However, a passage of the cooling frequency through the observational band could be an alternative scenario, as shown by the best-matching red curve.

The light curve of GRB\,061126 has a prominent bump in the 80 - 800\,s time interval \citep{perley2008}, which is excluded from the model. Otherwise, the early-time decay slope is too shallow to be successfully modeled with a RS. 

In the case of GRB\,080319B we model the  data after $\sim 100$ s. The steep decay of $\alpha \sim 6.5$ \citep{racusin2008} prior to this time cannot be explained within the RS model. 

We failed to model the light curve of GRB\,081007. The early-time peak ($t_{\rm p} \sim 130$ s), followed by a decay of $\alpha \sim 2$, can be explained with RS emission. However, the light curve then experiences a transition to a shallow decay phase with $\alpha \sim 0.65$, too shallow to be explained by a simple FS emission component. This shallow phase is explained by \citet{jin2013} as a FS emission with continued energy injection with an energy injection index $q = 0.5$ \citep{zhang2006}. This event is left out from further discussion.

The multi-wavelength light curve of GRB\,130427A has been found to be described well by a RS+FS thin-shell model with a wind environment  \citep{laskar2013,perley2013b}. However, excluding early reverse contributions, a FS model with ISM environment has also been successfully applied \citep{maselli2013}. In this work we only consider a constant density ISM circumburst environment. Our model can reproduce the optical light curve, assuming the break at $\approx 0.45$ days is a jet break. 

\subsection{Monte Carlo Simulation Results}
\label{MCresults}

\begin{deluxetable*}{lccccccccc}[H]  
\tablecolumns{7}
\tablecaption{Best-model parameters}
\tablehead{   
  \colhead{GRB} &
  \colhead{$p$} &
  \colhead{$\varepsilon_{\mathrm{e}}$ $[10^{-3}]$} &
  \colhead{$\varepsilon_{\mathrm{B,f}}$ [$10^{-5}$]} &
  \colhead{$R_{\mathrm{B}} =\varepsilon_{\mathrm{B,r}}/\varepsilon_{\mathrm{B,f}}$} &
  \colhead{$n$ [cm$^{-3}$]} &
  \colhead{$\Gamma_{0}$} &
  \colhead{$E_{\mathrm{K}}$ [$10^{52}$ erg]} &
  \colhead{$\eta_{\gamma}$} &
 \colhead{$\eta_{\gamma,\mathrm{max}}$}
}
\startdata
990123 & 2.49 & 79.0 & $5$ & 1156 &  $0.3$ & 420 & 108.0 & 0.2 $<$ $\eta_{\gamma}$ $<$ 0.9 & -\\ 
021004 & 2.57 & 260.0 &  $150$ & 5 &  $4.4$ & 99 & 7.0 & $<$ 0.8 & -\\
021211 & 2.20 & 130.0 &  $3$ & 128 &  $9.9$ & 154 & 3.0 & $<$ 0.6 & 0.1 \\
060908 & 2.24 & 14.0 &  $117$ & 72 &  $190.0$ & 107 & 2.7 & 0.5 $<$ $\eta_{\gamma}$ $<$ 0.9 & 0.7\\
061126 & 2.02 & 420.0 &  $8$ & 69 &  $3.7$ & 255 & 12.0 & 0.4 $<$ $\eta_{\gamma}$ $<$ 0.9 & 0.9\\
080319B & 2.57 & 68.0 &  $4$ & 16540 &  $0.6$ & 286 & 67.6 & $>$ 0.6 & $\sim 1$\\
090102 & 2.31 & 0.4 &  $2$ & 6666 &  $359.0$ & 228 & 816.0 & $<$ 0.4 & $<$ 0.1\\
090424 & 2.06 & 2.7 &  $19$ & 25 &  $4.0$ & 235 & 258.0 & $<$ 0.6 & 0.1\\
130427A & 2.08 & 3.3 & $22$ & 4 & $1.5$ & 157 & 521.0 & $<$ 0.8 & -
\enddata
\tablecomments{Parameters, corresponding to the models that provide the best match with the observational data (see red light curves in Figure \ref{fig8}). As discussed in Section \ref{MCresults}, the method we use does not allow us to constrain the parameters - they can only be constrained within the interval, shown by parameter distributions in Figure \ref{figtemp}. In the last two columns we report the spread of calculated radiative efficiency values and the most probable value of $\eta_{\gamma,\mathrm{max}}$ in the distribution.}
\label{tab3}
\end{deluxetable*}

\begin{figure*}[htp]
\centering
\begin{tabular}{cc}
\includegraphics[scale=0.45]{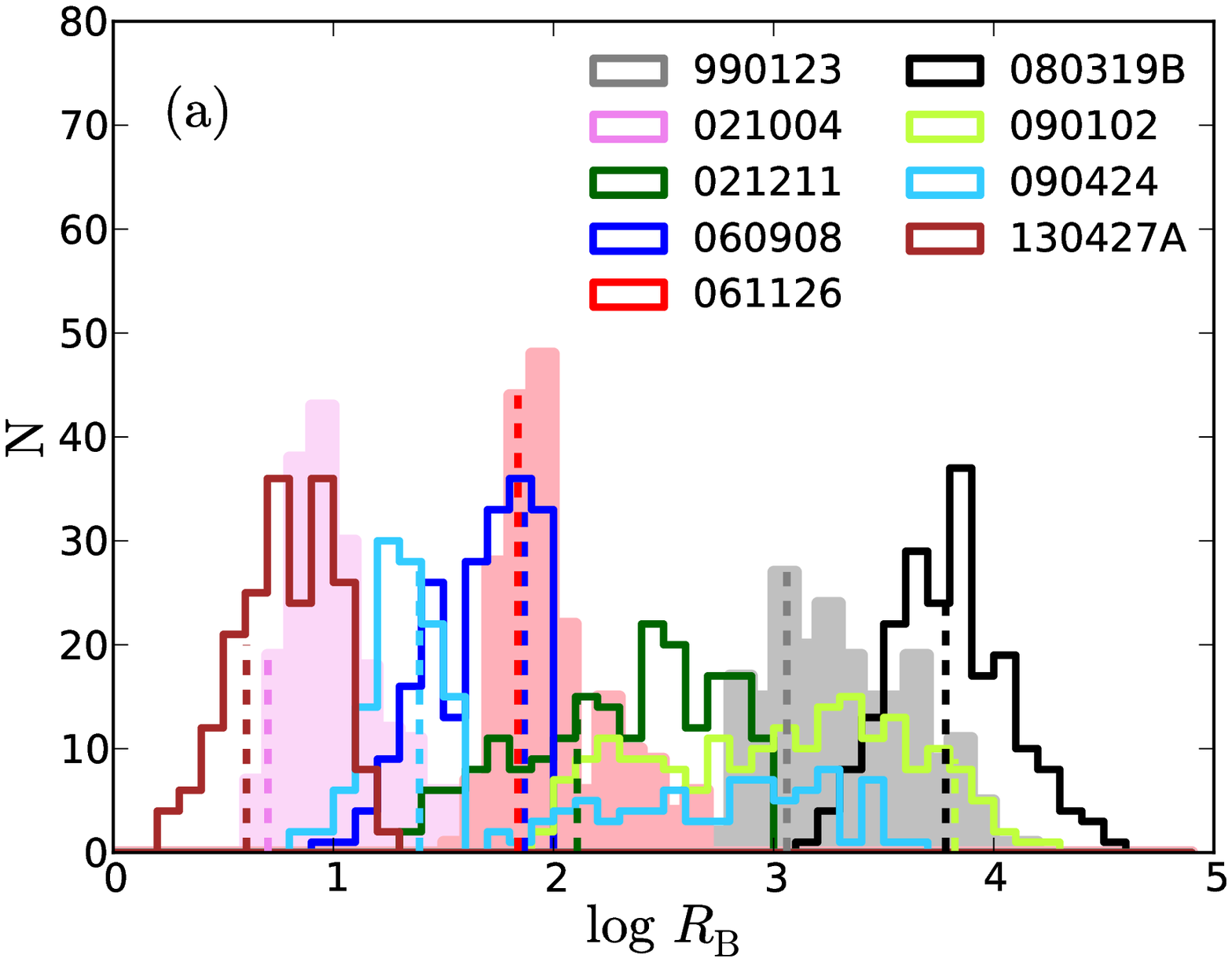}&
\includegraphics[scale=0.45]{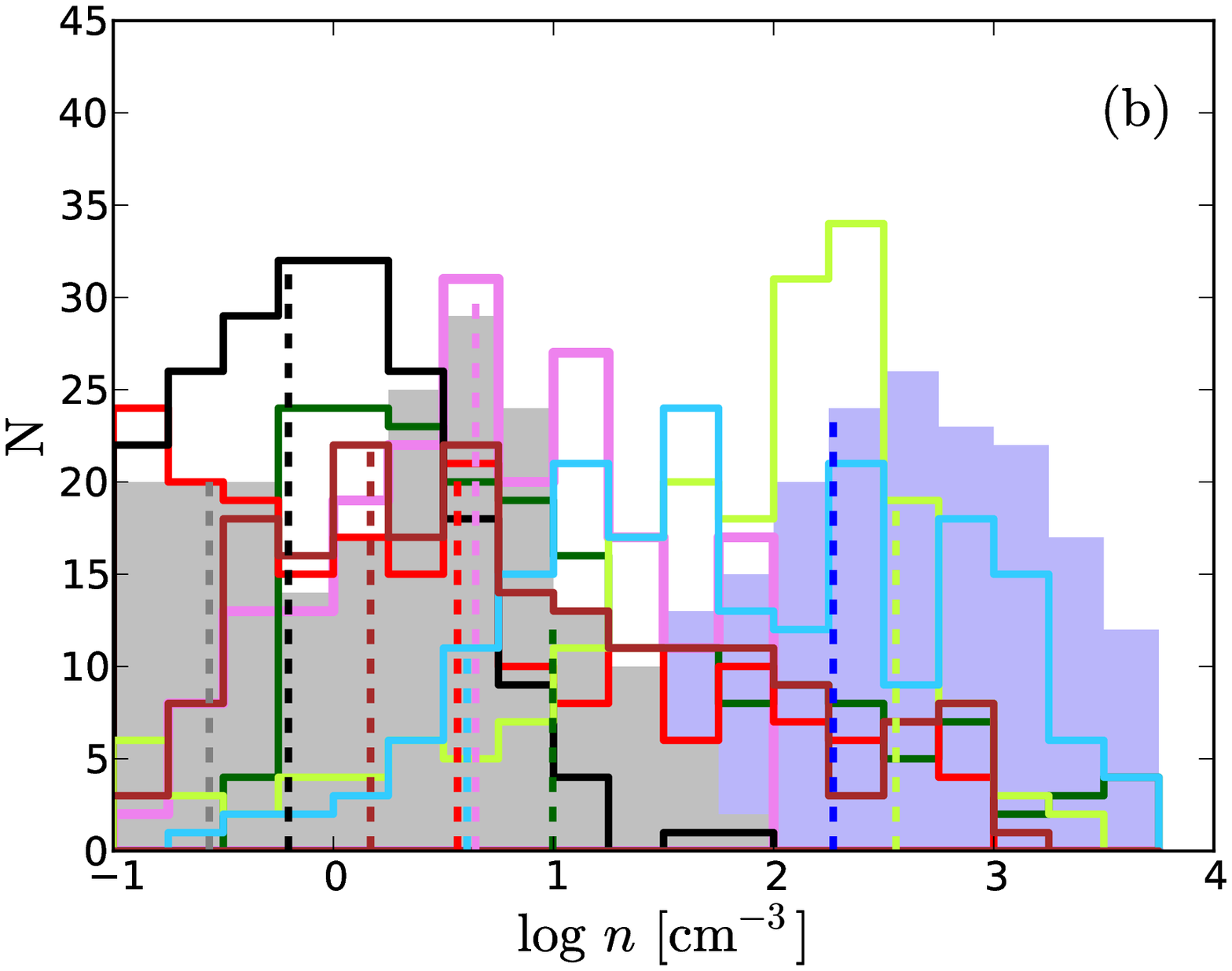}\\
\includegraphics[scale=0.45]{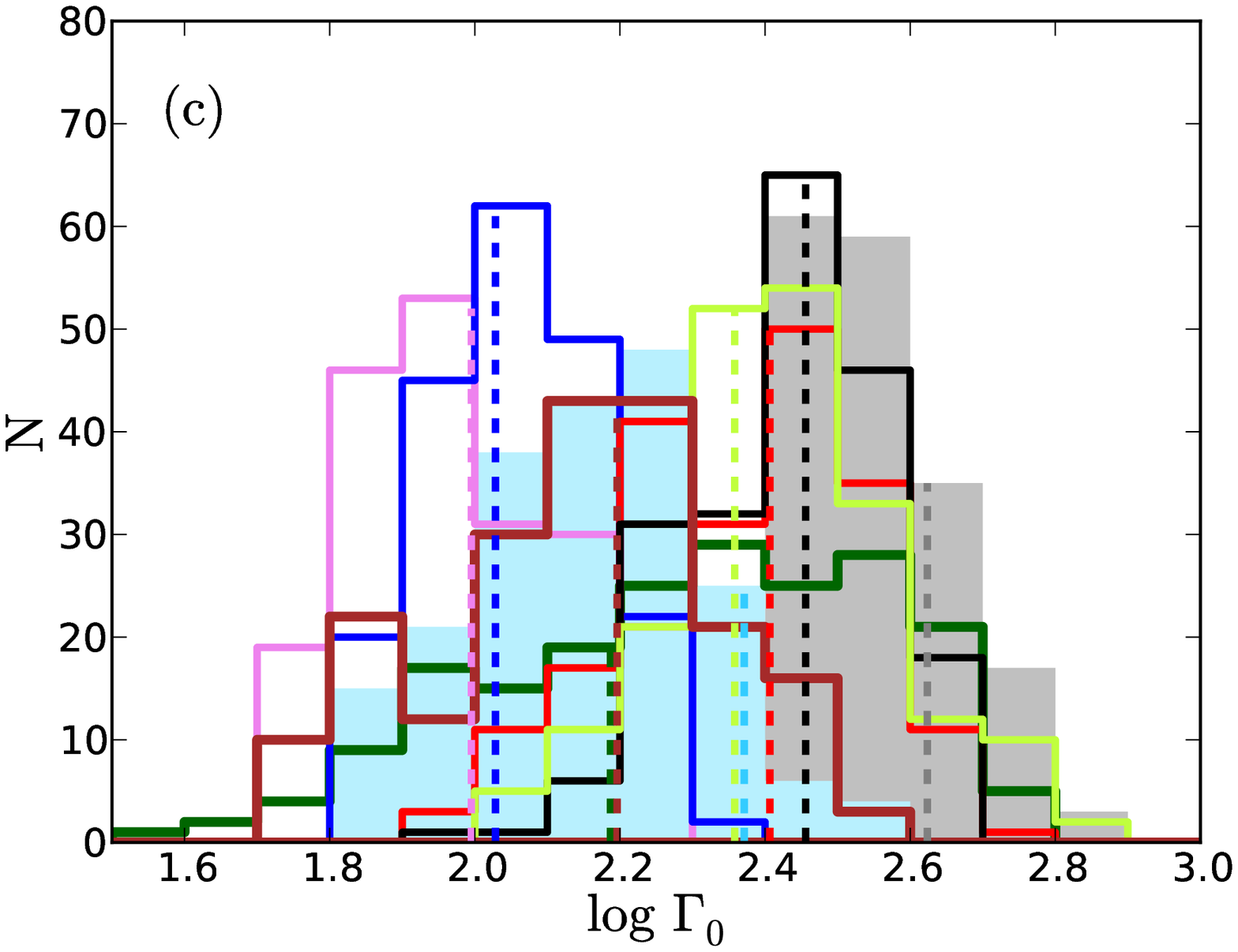}&
\includegraphics[scale=0.45]{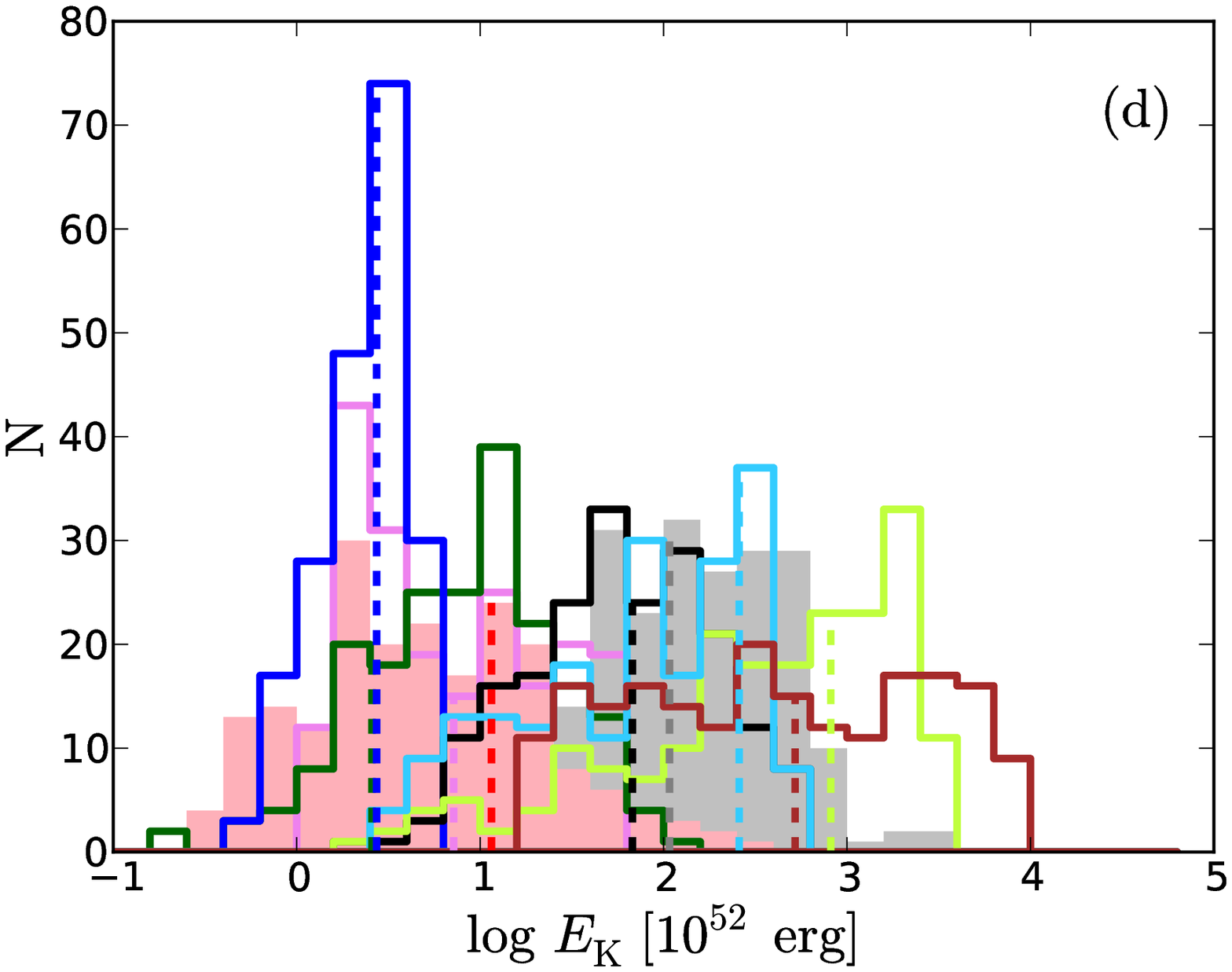}\\
\includegraphics[scale=0.45]{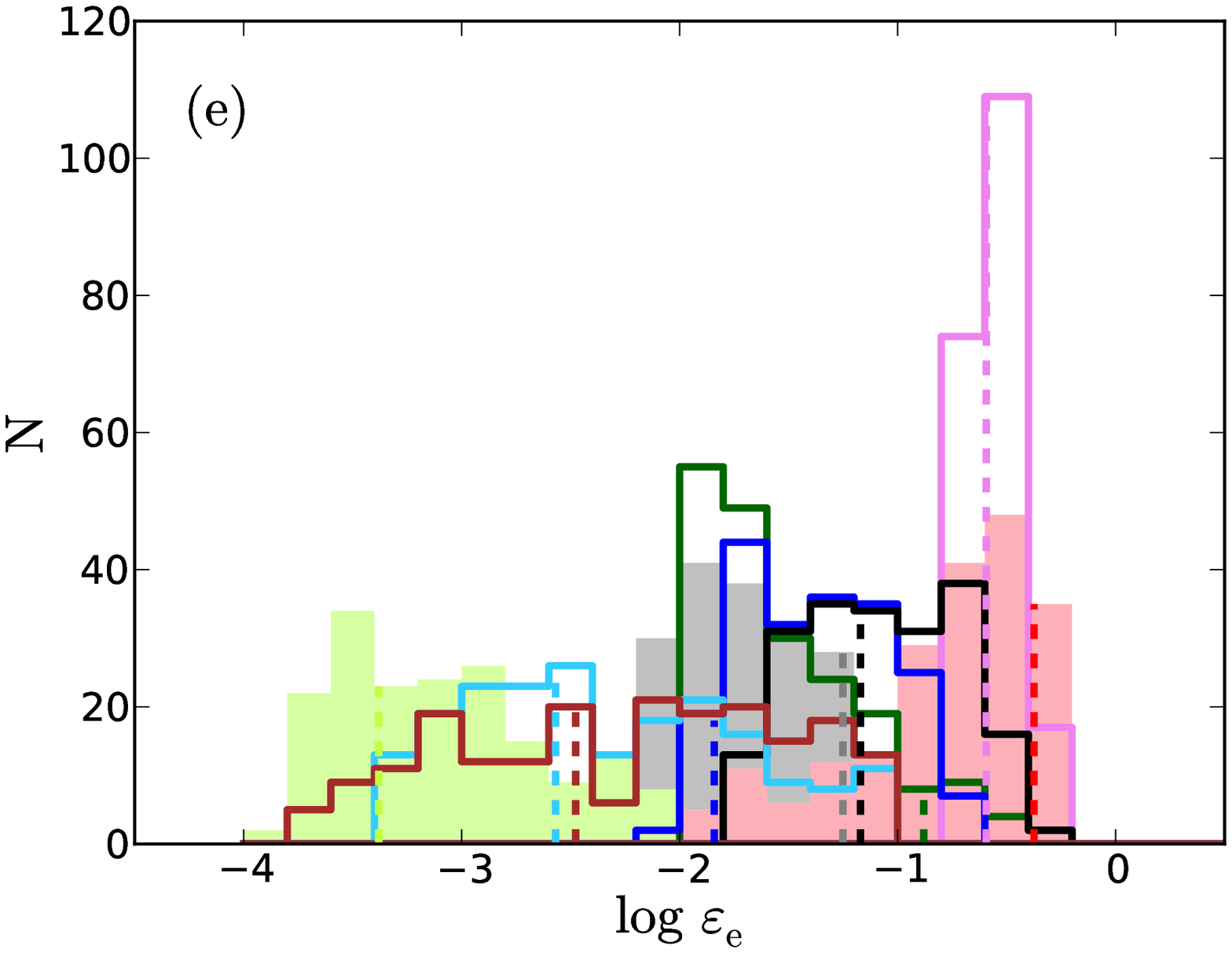}&
\includegraphics[scale=0.45]{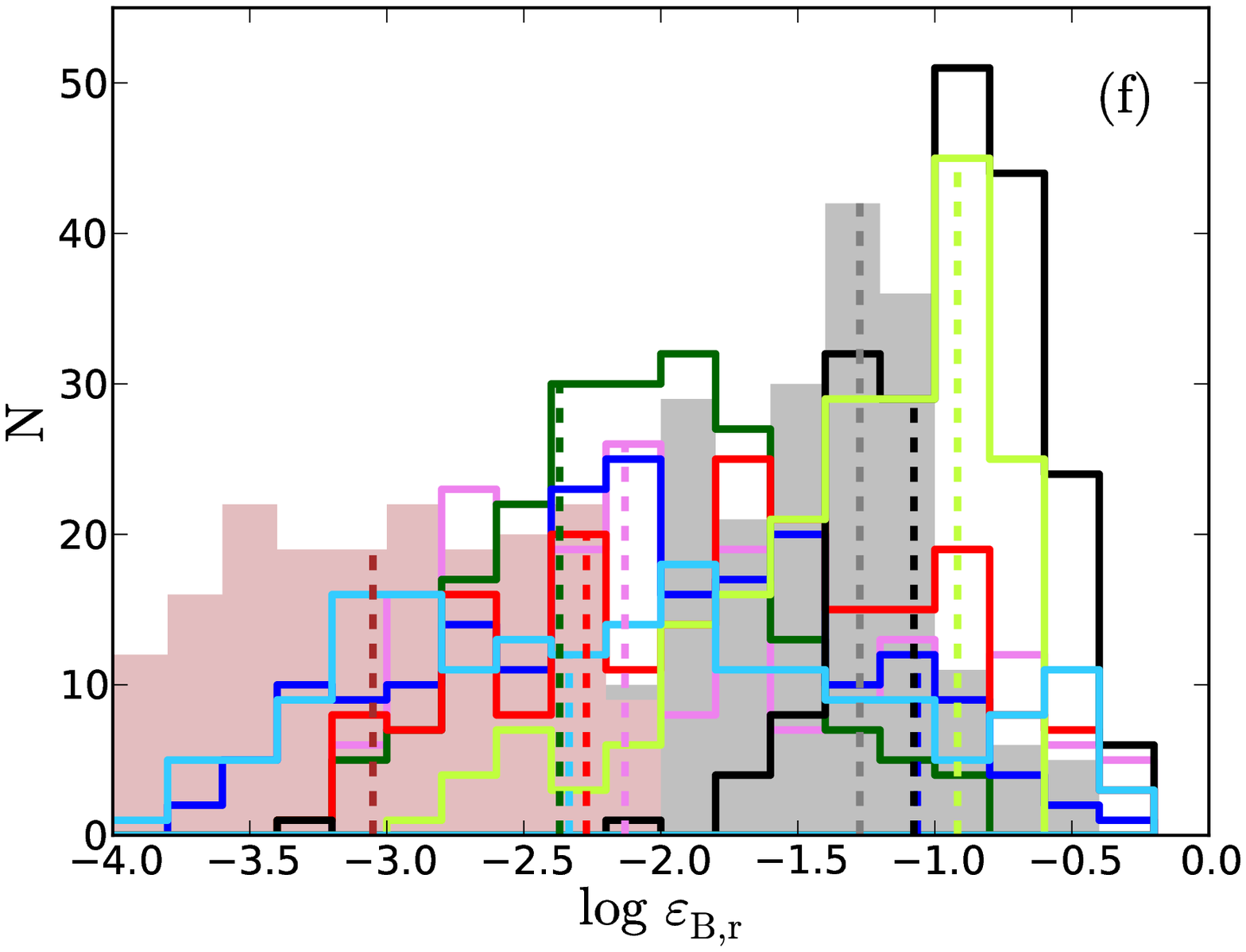}\\
 \end{tabular}
\caption{Distribution of parameter values $R_{\mathrm{B}}$, $n$, $\Gamma_{0}$, $E_{\mathrm{K}}$, $\varepsilon_{\mathrm{e}}$, and $\varepsilon_{\mathrm{B,r}}$ for 200 models that match the light curves well. Vertical dashed lines correspond to the positions of best-fit models: the values are provided in Table \ref{tab3}. Some histograms have been colored in order to make distributions easier to separate visually.}
 \label{figtemp}
\end{figure*}

In Figure \ref{figtemp} we show the distribution of parameters of the 200 generated light curves for each afterglow. From the results of the the simulations it appears that the physical conditions of our sample are very diverse: parameters occupy the whole predefined parameter space. 

The fractions of the kinetic energy deposited to a magnetic field in the reverse and forward shock region ($\varepsilon_{\mathrm{B,r}}$ and $\varepsilon_{\mathrm{B,f}}$) are unconstrained for most of the sample. The former, while generally low, spreads for about two orders of magnitude (or more) for each afterglow. Similarly,  $\varepsilon_{\mathrm{B,f}}$ occupies low values, spreading between $10^{-5} < \varepsilon_{\mathrm{B,f}} < 10^{-2}$. In the case of GRB 080319B the values are crowded toward the lower limit of  $10^{-5}$, suggesting even lower values are possible \citep[e.g., see ][]{panaitescu2004,santana2013}. 

The magnetization parameter $R_{\mathrm{B}}$ occupies values from $\approx$2 (GRBs 021004 and 130427A) to $\approx 10^4$ (GRB 080319B). Except for GRB\,090102, $R_{\mathrm{B}}$ is constrained within one order of magnitude for all cases. This is in contrast to the mostly unconstrained parameters $\varepsilon_{\mathrm{B,r}}$ and $\varepsilon_{\mathrm{B,f}}$. The reason for generally high $R_{\mathrm{B}}$ values is $\varepsilon_{\mathrm{B,r}}$, which has a role in the normalization of the RS afterglow. Most of our sample is composed of Type II light curves with a dominant RS component. Thus, high $\varepsilon_{\mathrm{B,r}}$ values relative to $\varepsilon_{\mathrm{B,f}}$ are needed to obtain strong RS afterglows and it is not a surprise to see such high values of $R_{\mathrm{B}}$ ratio. 

Values of the $\varepsilon_{\mathrm{e}}$ parameter are found in a wide range and are quite unconstrained for GRBs 061126, 090102, 090424 and 130427A. For  GRB\,090102, $\varepsilon_{\mathrm{e}}$ is especially low. Examining the 200 parameter sets for this GRB, we find that $\varepsilon_{\mathrm{e}}$ and $E_{\mathrm{K}}$ are strongly anti-correlated. Since the obtained $E_{\mathrm{K}}$ is also rather high, this suggests the preference toward low $\varepsilon_{\mathrm{e}}$ is a result of an unconstrained degeneracy between the two parameters. 

ISM density is found to be as high as $10^{4}$ cm$^{-3}$. The values for each GRB, obtained for different models, are spread over several orders of magnitude. The distributions for GRBs 990123, 061126 and 080319B imply that the densities could have values below the assumed lower limit of $10^{-1}$ cm$^{-3}$. However, removing the lower limit constraint, densities reach unrealistically low values down to $10^{-5}$ cm$^{-3}$. This is a consequence of degeneration between $n$, $E_{\rm K}$ and $\Gamma_{0}$: low values of the former result in high values of the latter two. For example, in the absence of the density constraint, $E_{\rm K}$ reaches values of $10^{55}$ ergs and more for GRBs 990123 and 080319B, implying very low prompt efficiency, which is unlikely for this two intrinsically very bright bursts.

Values of the initial Lorentz factor, $\Gamma_{0}$, lie between a few tens and $\approx 600$. This parameter is well constrained, which is a consequence of relation between $\Gamma_{0}$ and the deceleration time (the latter being at least partially constrained by the light curves): $t_{\mathrm{dec}} \propto E_{\mathrm{K}}^{1/3}n^{-1/3}\Gamma_{0}^{-8/3}$ \citep[e.g., ][]{kobayashi2000}. 

$E_{\mathrm{K}}$ occupies values between $10^{52}-10^{56}$ erg. Although it is not well constrained, we estimate the radiative efficiency of the prompt $\gamma$-ray emission, defined in \citet{zhang2007} as $\eta_{\gamma} = E_{\gamma,\mathrm{iso}}/(E_{\gamma,\mathrm{iso}} + E_{\mathrm{K}})$. Assuming the $E_{\gamma,\mathrm{iso}}$ values given in Table \ref{tab1}, we obtain an efficiency for each burst, reported in Table \ref{tab3}.  The efficiency is high for GRBs 060908, 061126 and 080319B, low for GRBs, 021211, 090102 and 090424 and mostly unconstrained for GRBs 990123, 021004 and 130427A. The latter two have most of the values very near the limiting efficiency of $\eta_{\gamma} = 0.01$, suggesting either the efficiency is even lower than that or, more likely, the degeneracy of $E_{\mathrm{K}}$ with other parameters is affecting our results. A large spread in derived most-probable efficiency values $\eta_{\gamma, \mathrm{max}}$ (that is, $\eta_{\gamma}$ at the peak of $E_{\mathrm{K}}$ distribution) is in agreement with the values derived in \citet{zhang2007}. 

The spread in parameter values varies considerably. In some cases the values are completely unconstrained and occupy several orders of parameter space. The spread is an expected consequence of using monochromatic observations to constrain the analytic model: as already mentioned parameters can be severely degenerated. For example, parameters of the best and second best model (according to $\chi^2$ statistics) can differ for an order of magnitude. This degeneration is the reason why some of the parameters corresponding to the best matching light curve do not represent the peak of the distributions, plotted in Figure \ref{figtemp}. Taking a subsample of models within the 200 light curves that best match observations (i.e., using $\chi^2$ statistics) does not reduce the spread in the distributions. Our analysis shows that monochromatic light curves are not enough to constrain the parameters much better than shown by the widths of the distributions. Due to the nature of our MC simulation, a small fraction of parameter sets results in light curves that visually (and statistically) do not match the data very well: these cases have very high $\chi^2$ values in the tail of $\chi^2$ distribution. The parameters corresponding to these cases are not constrained to one region of their parameter space, but they do occur preferentially in the tails of their distributions.

Comparing our results with previous analyses is not trivial: while most of the models are based on the standard fireball model, different studies use different emission components or circumburts environment properties in order to explain observations. \citet{harrison2013} undertook a numerical approach to describe conditions in the intermediate RS shell regime between relativistic and sub-relativistic. They specifically calculate the $R_{\mathrm{B}}$ ratios for GRBs 990123 and 090102. Our results agree well with theirs (though, admittedly, our $R_{\mathrm{B}}$ parameter in the case of GRB\,090102 is mostly unconstrained). \citet{panaitescu2004} constructed several different models, trying to explain the afterglows of GRBs 990123 and 021211. They found that a constant ISM thick-shell RS + FS case with highly radiative dynamics can describe the observed optical and radio light curves of GRB\,990123, with $R_{\mathrm{B}} > 100$, $E_{\mathrm{K}} \gtrsim 10^{55}$ erg and $n \gtrsim 1$ cm$^{-3}$: the latter is in complete contrast to our result. Their model, however, assumed that $\varepsilon_{\mathrm{e}}$ as well as $\varepsilon_{\mathrm{B}}$ differ in front of and behind the contact discontinuity. They could also reproduce the light curve with a fully radiative model in a wind environment but with the same microphysical parameters in the two regions. A similar analysis was done for GRB\,021211, where $R_{\mathrm{B}}$ was found to be either very high (thin-shell; $R_{\mathrm{B}} > 10^3$) or very low (thick-shell; $R_{\mathrm{B}} \sim 1$). Instead, we find $R_{\mathrm{B}}$ to lie in between these two values. Our derived values of $R_{\mathrm{B}}$ for GRB\,061126 are in agreement with the one given by \citet{gomboc2008} ($R_{\mathrm{B}} \sim 50$). The results we obtain for GRB\,021004 agree with \citet{kobayashi2003}. The parameter estimates we obtain for GRB\,130427A differ from the values estimated by \citet{perley2013b}. This is not surprising, since we use a different theoretical premise for modeling (i.e., wind versus ISM circumburst medium). The exception is $R_{\rm B}$, which we found to be low in both studies.

Overall, while the parameter values themselves have a large spread, it seems that very diverse physical properties can be found in the GRB ejecta and their surrounding environment. Similar results, using different models, have been obtained in previous studies.  By modeling several afterglows with a FS model, Panaitescu \& Kumar (2001; 2002) found that $n$ and $\varepsilon_{\mathrm{B,f}}$ values occupy similarly wide parameter spaces as found in this work. Recently, \citet{santana2013} modeled FS afterglow emission and, assuming a constant $\varepsilon_{\rm e} = 0.2$, $n = 1$ cm$^{-3}$ and $\eta = 0.2$, found low values of $\varepsilon_{\rm B,f}$ ($10^{-8} - 10^{-3}$), which are generally lower than values obtained in this and previous studies. 

According to our findings, most of the cases in our sample are allowed rather low values of $\varepsilon_{\rm e}$ and $\varepsilon_{\rm B,r}$. If the prompt emission is due to internal shock scenario, the same microphysical parameters should in principle be used both for internal shocks and external reverse shocks. The small values we infer from the modeling suggest an inefficient gamma-ray burst, peaking at low energies. This inconsistency with the observed data (and rough efficiency values reported in Table \ref{tab3}) may be resolved if the jetted outflow is highly variable, i.e., composed of shells with Lorentz factors differing by a few orders of magnitude \citep[e.g.,][]{kobayashi2001}. In this case, the density of different shells should not vary too much, otherwise its irregularity could survive the internal shock phase and affect the reverse shock evolution (and thus our initial assumptions). We also note that $\varepsilon_{\rm e}$ has been chosen to be the same in the reverse and forward shock region while in principle this is not necessarily true.

\subsection{Radio afterglows}
\label{radio}

We check whether the models that provide a good match in optical wavelengths can also reproduce observations at other energies. Five GRBs in the sample (GRBs 990123, 021004, 080319B, 090424 and 130427A) were detected in radio wavelengths. To calculate the radio afterglow, we modify the code in order to account for the synchrotron self-absorption effect. We calculate the value of the self-absorption frequency, $\nu_{\mathrm{sa,f}}$, assuming the expression given by \citet{granot2002}\footnote{The difference between light curves obtained by the model we used and the ones obtained by using the model given by \citet{granot2002} is small, compared to differences with some other models \citep{granot2002}. There is practically no difference in calculated $\nu_{\mathrm{m,f}}$ and $\nu_{\mathrm{c,f}}$ ($< 4\%$), while the absolute flux differs for a factor of  $\sim 1.7-2.0$. Assuming $\nu_{\mathrm{sa}}$ provided by \citet{granot2002} should not considerably affect our results.} and take into account the fact that the flux density below $\nu_{\mathrm{sa,f}}$ drops significantly $(F_{\nu} \propto \nu^{2})$. In the case of a RS afterglow, we assume a simple estimate for the upper-limit of the self-absorbed flux to be an emission from a  black body with the RS temperature \citep{sari1999b,kobayashi2000b, melandri2010}. The flux density of a black body at the deceleration time is \citep[see e.g., ][Eq. 6]{melandri2010}
\begin{eqnarray}\nonumber
F_{\nu,\mathrm{BB}} &\approx & 1.3\times 10^{-14} (1 + z)\times \\
&& \times \varepsilon_{\mathrm{e}}\Gamma_{0}^{3}\nu_{9}^{2}D_{\mathrm{L},28}^{-2}\left(\frac{t_{\mathrm{dec}}}{\mathrm{s}}\right)^{2}~\mathrm{mJy}.
\end{eqnarray}
This limit initially increases $\propto t^{1/2}$. After the typical frequency, $\nu_{\mathrm{m,r}}$, crosses the observed band, the increase steepens $(\propto t^{5/4})$. The combination of the increasing limit and decaying RS emission produces a flare \citep[for a schematic plot of the emission components, see Figure 8 in ][]{melandri2010}.

Applying the above prescription to our model, we calculated the radio light curves, assuming previously obtained parameter sets. In general, we find the radio flux is overestimated by the models by a factor of $\approx 5 - 10$ in the best of cases. Repeating the simulation with an additional radio constraint for these GRBs, we do not find a parameter space in which the optical and radio flux can be simultaneously reproduced with the assumed model within reasonable accuracy. Our inability to reproduce the radio observations may be due to the assumption of a thin- rather than thick-shell RS evolution. A homogeneous environment may also be too simple an approximation, and GRB\,130427A, at least, has been successfully modeled by a wind environment \citep{laskar2013,perley2013b}. On the other hand, \citet{liang2013} and \citet{yi2013} recently investigated temporal evolution and resulting emission of RS and FS for a general environment density distribution (e.g., $n \propto r^{-k}$) and found that the environment is neither purely homogeneous nor stellar wind. Alternatively, assuming that all the microphysical parameters have different values in front of and behind the contact discontinuity might help to reproduce other wavebands \citep{panaitescu2004}.

GRB\,021004 can in principle be reproduced at radio wavelengths, but a more complete light curve in mm wavelengths \citep{postigo2005} reveals our models overpredict the peak time in the mm band by a factor of $\sim 10$. This result suggests that the peak in optical we assumed to be at $t_{\mathrm{p}} \sim 0.1$ days might not be due to passage of $\nu_{\mathrm{m,f}}$ through the optical band but rather an energy injection, as claimed by \citet{postigo2005}. 

\subsection{X-ray afterglows}

All seven \textit{Swift}-era afterglows from our RS sample have also been observed with the XRT instrument onboard \textit{Swift}. We check whether the models that match the optical data well can also reproduce the X-ray afterglows. For this we calculate the corresponding light curves at 10 keV and compare them to the unabsorbed flux density light curves available in the online light curve repository\footnote{http://www.swift.ac.uk/burst$\_$analyser/} \citep{evans2010}. 

X-ray light curves are reproduced well only for GRB\,090424. The X-ray flux is underestimated by  a factor of $\sim 10$ in the case of GRB\,090102. For all other cases (except GRB\,081007, for which we can model neither the optical nor X-ray band) we obtain a correct absolute flux scale but incorrect decay slopes. The difference between the decay indices of optical and X-ray light curves at late times (in a constant density ISM medium) is expected to be $\Delta \alpha = 0$ if both observational bands are in the same spectral regime, or $\Delta \alpha = 0.25$ after the passage of cooling frequency between optical and X-ray band \citep{sari1998}. However, using a large sample of afterglows with optical and X-ray observations, \citet{zaninoni2013} found that only $20\%$ of the events follow this theoretical prediction. We compare decay indices of late-time optical and X-ray light curves and find the following slope differences: $\Delta \alpha = 0.66 \pm 0.06,~0.59 \pm 0.08,~0.62 \pm 0.15$, $0.68 \pm 0.07$, $0.26 \pm 0.06$ and $0.47 \pm 0.05$  for GRB\,060908, 061126, 080319B, 090102, 090424 and 130427A, respectively. Due to differences in decay indices we cannot use the X-ray light curves to further constrain our results.

\subsection{Modeling caveats}
\label{caveats}
The simple model we use has several caveats. \citet{granot2002} have demonstrated that different variations of the standard FS model, when used to derive values of physical parameters, give results that in extreme cases may differ by several orders of magnitude. The model does not incorporate emission produced by the inverse Compton effect, which is expected to delay the transition between the fast- and slow-cooling phase and decrease the cooling frequency in the FS \citep{wu2005}. The latter could affect our results only in the case of GRBs 021004 and 060908, since in all other GRBs the cooling frequency is much higher than the optical band at the times of our analysis. The effect, whose contribution is non-negligible in dense environments with $n > 1 \mathrm{cm}^{-3}$ \citep{sari2001,wu2005}, is especially important at high energies and may contribute to the failure of this models to reproduce the X-ray light curves (see previous discussion). In principle, early FS evolution should be modified to include radiation losses - a correction that is dependent on $\varepsilon_{\mathrm{e}}$ \citep{sari1997}. We also model sharp light curve breaks whereas in reality these breaks are smooth \citep{granot2002}. We note that we only consider afterglows produced in an ISM environment. Wind environment models have different parameter dependencies and different light curve evolution \citep{chevalier2000,kobayashi2003b,zou2005}. In addition, RS time evolution and its dependency on the various parameters differs for thin- and thick-shell models.

\section{Conclusions}
In this work we present a detailed study of a sample of 10 GRB afterglows that show RS signatures in early optical light curves. The sample is composed of one Type I (in which both reverse and forward shock afterglow light curve peaks are observed) and  nine Type II light curves (in which the characteristic steep-to-shallow light curve evolution, caused by the dominant RS at early time and the later rise of FS emission, is observed), as classified in \citet{zhang2003} and \citet{gomboc2009}. The 10 afterglows represent only a fraction of a much larger sample, composed of 118 afterglows with measured redshift and host galaxy extinction, which we compiled in order to investigate the rest-frame properties of the former in relation to the larger sample. 

We compare the rest-frame optical, X-ray and $\gamma$-ray properties of the RS sample to a sample of afterglows without compelling RS signatures. Early-time RS emission is found to span over several orders in spectral luminosity, which is consistent with the general early-time spread in afterglows' brightness \citep[e.g., ][]{kann2010}. On the other hand, we find that all but one afterglow from our sample are among the faintest at late times ($t_{\mathrm{rest}} > 10$ ks). Since only 6 out of 10 RS afterglows were observed with the \textit{Swift} XRT instrument, we cannot draw any firm conclusions about the X-ray properties of the sample. The high-energy properties (i.e., isotropic equivalent energy) of RS and non-RS GRBs in our full sample do not statistically differ.

Using a simple analytic model of a RS and FS afterglow we reproduce the observed optical light curves of our RS sample by using a MC simulation. Derived physical properties do not reveal any preferential values within the assumed parameter space. This is similar to the results obtained in previous analyses which concentrated on late time FS emission \citep[e.g., ][]{panaitescu2001,panaitescu2002}. Failure to reproduce X-ray and radio observations, where available, points to the need to either change the basic assumptions of the model (e.g., thick- vs. thin-shell scenario, ISM vs. stellar wind circumburst medium) or introduce more complicated emission components beyond the simple standard theory.

According to \citet{zhang2005}, a strong RS emission is produced when the GRB outflow is baryonic, i.e., only mildly magnetized. Furthermore, in order to produce a RS afterglow that can outshine the FS emission (Type II light curve), a magnetization parameter of $R_{\rm B} > 1$ is required. We find that our RS sample afterglows have preferentially both low $\varepsilon_{\rm B,r}$ as well as high $R_{\rm B}$ values. Consequently,  the presence of strong RS emission (compared to  FS emission) requires $\varepsilon_{\rm B,f} < \varepsilon_{\rm B,r}$. In the standard FS afterglow model, the peak in the spectral domain $F_{\rm max,f}$ is proportional to $\varepsilon_{\rm B,f}^{1/2}$ \citep{sari1998}. Thus, a low value of $\varepsilon_{\rm B,f}$ is expected to produce fainter FS emission, which is what we find in our RS sample at optical wavelengths. In addition, the time of the FS peak is proportional to $t_{\rm p} \propto \varepsilon_{\rm B,f}^{1/3}$ \citep{sari1998}. This could explain the lack of Type I light curves, since the FS peak for low $R_{\mathrm{B}}$ ratio is likely to occur when the RS afterglow component is still very bright. Due to different models used in the literature as well as our mostly unconstrained values of $\varepsilon_{\rm B,f}$, we cannot test whether the derived $\varepsilon_{\rm B,f}$ in afterglows with prominent RS components is generally lower than in non-RS events. The interpretation of faint late-time optical afterglows in our RS sample may be revised if there is an intrinsic correlation between $\varepsilon_{\rm B,f}$ and other parameters that define the afterglow emission (e.g., \citealt{santana2013} recently found a hint of correlation between parameters $\varepsilon_{\rm B,f}$ and $E_{\rm K}$).  

Fifteen years after the discovery of GRB\,990123, it is clear that larger samples of confirmed RS components are vital to understand the nature of RS emission and to determine the origin of RS suppression. In addition to the standard techniques involving light curve and spectral analysis, unambiguous identification of RS components may become increasingly possible via the detection of early time optical polarization \citep{mundell2007a,steele2009,mundell2013} with simultaneous multicolor light curves using new polarimeters on robotic telescopes, such as RINGO3, mounted on the Liverpool Telescope \citep{arnold2012}.

\acknowledgments
We thank the anonymous referee for useful comments and suggestions. This work made use of data supplied by the UK Swift Science Data Centre at the University of Leicester. J. J., A.G., and D. K. acknowledge funding from the Slovenian Research Agency. CGM thanks the Royal Society, the Wolfson Foundation and the Science and Technology Facilities Council (STFC) for funding.


\appendix

\setcounter{table}{0} \renewcommand{\thetable}{A.\arabic{table}}

\section{Afterglow light curve data references}
\textbf{GRB 970508:} \citet{bloom1998} \citet{castro1998} \citet{djorgovski1997} \citet{galama1998} \citet{garcia1998} \citet{kelemen1997} 
\citet{pedersen1998} \citet{pian1998} \citet{sokolov1998} \citet{zharikov1998}
\textbf{GRB 971214:} \citet{diercks1998} \citet{halpern1998} \citet{itoh1997} \citet{kulkarni1998} \citet{sokolov2001}
\textbf{GRB 980703:} \citet{bloom1998b} \citet{vreeswijk1999} \citet{castro1999} \citet{holland2001} \citet{sokolov2001}
\textbf{GRB 990123:} \citet{castro1999b} \citet{kulkarni1999} \citet{akerlof1999} \citet{fruchter1999} \citet{galama1999} \citet{holland1999} \citet{sokolov2001} \citet{halpern1999} \citet{maury1999} \citet{sagar1999}
\textbf{GRB 990510:} \citet{harrison1999} \citet{israel1999} \citet{beuermann1999} \citet{galama1999b}
\textbf{GRB 991208:} \citet{castro2001} \citet{dodonov1999} \citet{garnavich1999}
\textbf{GRB 991216:} \citet{garnavich2000} \citet{halpern2000}
\textbf{GRB 000301C:} \citet{jensen2001} \citet{bhargavi2000} \citet{masetti2000} \citet{rhoads2001} \citet{sagar2000} \citet{sokolov2001}
\textbf{GRB 000418:} \citet{klose2000} \citet{berger2001} 
\textbf{GRB 000911:} \citet{lazzati2001} \citet{masetti2005} \citet{price2002} 
\textbf{GRB 000926:} \citet{fynbo2001} \citet{price2001} \citet{harrison2001}
\textbf{GRB 010222:} \citet{galama2003} \citet{masetti2001} \citet{cowsik2001} \citet{oksanen2002} \citet{sagar2001} \citet{stanek2001} \citet{watanabe2001}
\textbf{GRB 010921:} \citet{price2002b} \citet{park2002}
\textbf{GRB 011121:} \citet{garnavich2003} \citet{bloom2002} \citet{greiner2003}
\textbf{GRB 011211:} \citet{holland2002} \citet{jakobsson2003} \citet{jakobsson2004}
\textbf{GRB 020124:} \citet{hjorth2003} \citet{berger2002}
\textbf{GRB 020813:} \citet{covino2003} \citet{li2003} \citet{urata2003}
\textbf{GRB 021004:} \citet{holland2003} \citet{uemura2003} \citet{pandey2003} \citet{fox2002}
\textbf{GRB 021211:} \citet{holland2004} \citet{pandey2003b} \citet{li2003b} \citet{fox2003}
\textbf{GRB 030226:} \citet{klose2004} \citet{pandey2004} \citet{ando2003} \citet{garnavich2003b} \citet{vonbraun2003} \citet{rumyantsev2003} \citet{covino2003b} \citet{fatkhullin2003} \citet{rumyantsev2003b}
\textbf{GRB 030227:} \citet{castro2003}
\textbf{GRB 030328:} \citet{maiorano2006}
\textbf{GRB 030329:} \citet{price2003} \citet{lipkin2004} \citet{resmi2005} \citet{torii2003}
\textbf{GRB 030429:} \citet{jakobsson2004b} \citet{rumyantsev2003c}
\textbf{GRB 040924:} \citet{wiersema2008} \citet{soderberg2006} \citet{huang2005}
\textbf{GRB 041006:} \citet{stanek2005} \citet{urata2007} \citet{melandri2008}
\textbf{GRB 041219A:} \citet{blake2005}
\textbf{GRB 050318:} \citet{still2005}
\textbf{GRB 050319:} \citet{wozniak2005} \citet{quimby2006} \citet{huang2007} \citet{kamble2007}
\textbf{GRB 050401:} \citet{watson2006} \citet{rykoff2005} \citet{pasquale2006}
\textbf{GRB 050408:} \citet{postigo2007} \citet{foley2006} \citet{wiersema2005} \citet{milne2005}
\textbf{GRB 050416A:} \citet{soderberg2007} \citet{holland2007}
\textbf{GRB 050502A:} \citet{guidorzi2005} \citet{yost2006}
\textbf{GRB 050525A:} \citet{blustin2006} \citet{klotz2005} \citet{dellavalle2006} \citet{rykoff2009}
\textbf{GRB 050730:} \citet{pandey2006}
\textbf{GRB 050801:} \citet{pasquale2007} \citet{rykoff2006} \citet{fynbo2005}
\textbf{GRB 050802:} \citet{oates2007}
\textbf{GRB 050820A:} \citet{cenko2006} \citet{vestrand2006}
\textbf{GRB 050824:} \citet{sollerman2007} \citet{lipunov2005}
\textbf{GRB 050904:} \citet{tagliaferri2005} \citet{boer2006} \citet{berger2007}
\textbf{GRB 050922C:} \citet{rykoff2009} \citet{durig2005}
\textbf{GRB 051109A:} \citet{yost2007} \citet{huang2005b} \citet{li2005} \citet{misra2005} \citet{kinugasa2005}
\textbf{GRB 051111:} \citet{butler2006} \citet{guidorzi2007} \citet{yost2007}
\textbf{GRB 060124:} \citet{misra2007} \citet{covino2006}
\textbf{GRB 060206:} \citet{monfardini2006} \citet{wozniak2006} \citet{stanek2007}
\textbf{GRB 060210:} \citet{curran2007} \citet{stanek2007} \citet{cenko2009}
\textbf{GRB 060418:} \citet{molinari2007} \citet{melandri2008} \citet{cenko2010a}
\textbf{GRB 060502A:} \citet{cenko2009} \citet{jakobsson2006b}
\textbf{GRB 060512:} \citet{melandri2008}
\textbf{GRB 060526:} \citet{thone2010}
\textbf{GRB 060605:} \citet{ferrero2009}
\textbf{GRB 060607A:} \citet{nysewander2009} \citet{molinari2007}
\textbf{GRB 060614:} \citet{mangano2007} \citet{dellavalle2006b} \citet{fynbo2006b} \citet{xu2009}
\textbf{GRB 060729:} \citet{grupe2007}
\textbf{GRB 060904B:} \citet{klotz2008} \citet{rykoff2009}
\textbf{GRB 060908:} \citet{covino2010} \citet{cenko2009}
\textbf{GRB 060912A:} \citet{deng}
\textbf{GRB 060927:} \citet{ruizvelasco2007}
\textbf{GRB 061007:} \citet{mundell2007b} \citet{rykoff2009}
\textbf{GRB 061121:} \citet{page2007}
\textbf{GRB 061126:} \citet{gomboc2008} \citet{perley2008}
\textbf{GRB 070125:} \citet{updike2008} \citet{chandra2008} \citet{dai2008} \citet{yoshida2007} \citet{uemura2007}
\textbf{GRB 070208:} \citet{melandri2008} \citet{cenko2009} \citet{swan2007} \citet{klunko2007}
\textbf{GRB 070306:} \citet{jaunsen2008}
\textbf{GRB 070419A:} \citet{melandri2009} \citet{cenko2009} \citet{dai2008}
\textbf{GRB 070802:} \citet{kruhler2008} \citet{eliasdottir2009}
\textbf{GRB 071003:} \citet{perley2008} \citet{guidorzi2007b}
\textbf{GRB 071010A:} \citet{covino2008} \citet{cenko2009}
\textbf{GRB 071020:} \citet{cenko2009} \citet{schaefer2007} \citet{updike2007} \citet{lee2010} \citet{xin2007}
\textbf{GRB 071025:} \citet{perley2010}
\textbf{GRB 071031:} \citet{kruhler2009}
\textbf{GRB 071112C:} \citet{huang2012}
\textbf{GRB 080129:} \citet{greiner2009} \citet{perley2008c}
\textbf{GRB 080310:} \citet{cenko2009} \citet{littlejohns2012}
\textbf{GRB 080319B:} \citet{bloom2009} \citet{racusin2008} \citet{cenko2009} \citet{wozniak2009} \citet{tanvir2010} \citet{pandey2009} \citet{karpov2008}
\textbf{GRB 080319C:} \citet{cenko2009} \citet{williams2008} \citet{li2008} \citet{wren2008}
\textbf{GRB 080330:} \citet{guidorzi2009} \citet{wren2008b}
\textbf{GRB 080413B:} \citet{filgas2011}
\textbf{GRB 080603A:} \citet{guidorzi2011}
\textbf{GRB 080603B:}  \citet{jelinek2012} \citet{ibrahimov2008} \citet{rumyantsev2008} \citet{klunko2008} \citet{xin2008}
\textbf{GRB 080607:} \citet{perley2011}
\textbf{GRB 080710:} \citet{kruhler2009b} \citet{yoshida2008}
\textbf{GRB 080721:} \citet{starling2009}
\textbf{GRB 080810:} \citet{page2009}
\textbf{GRB 080913:} \citet{greiner2009b}
\textbf{GRB 080916A:} \citet{covino2013}
\textbf{GRB 080916C:} \citet{greiner2009c}
\textbf{GRB 080928:} \citet{rossi2011} \citet{ferrero2008}
\textbf{GRB 081007:} \citet{jin2013}
\textbf{GRB 081008:} \citet{yuan2010}
\textbf{GRB 081029:} \citet{nardini2011}
\textbf{GRB 081203A:} \citet{kuin2009} \citet{andreev2008} \citet{volkov2008} \citet{west2008} \citet{mori2008} \citet{isogai2008} \citet{rumyantsev2008b}
\textbf{GRB 081222:} \citet{covino2013}
\textbf{GRB 090102:} \citet{gendre2010} \citet{postigo2009b} \citet{cenko2009b}
\textbf{GRB 090313:} \citet{melandri2010}
\textbf{GRB 090323:} \citet{mcbreen2010} \citet{cenko2011}
\textbf{GRB 090328:} \citet{mcbreen2010}
\textbf{GRB 090423:} \citet{tanvir2009} \citet{yoshida2009}
\textbf{GRB 090424:} \citet{jin2013}
\textbf{GRB 090618:} \citet{cano2011}
\textbf{GRB 090709A:} \citet{cenko2010b}
\textbf{GRB 090902B:} \citet{pandey2010} \citet{cenko2011} \citet{mcbreen2010}
\textbf{GRB 090926A:} \citet{cenko2011} \citet{rau2010}
\textbf{GRB 091018:} \citet{wiersema2012}
\textbf{GRB 091029:} \citet{filgas2012}
\textbf{GRB 091127:} \citet{troja2012} \citet{cobb2010}
\textbf{GRB 091208B:} \citet{uehara2012}
\textbf{GRB 100219A:} \citet{thoene2012} \citet{mao2012}
\textbf{GRB 100418A:} \citet{marshall2011}
\textbf{GRB 100621A:} \citet{kruhler2011}
\textbf{GRB 100901A:} \citet{gomboc2013}
\textbf{GRB 110205A:} \citet{gendre2012} \citet{cucchiara2011} \citet{zheng2012}
\textbf{GRB 110731A:} \citet{ackermann2013}
\textbf{GRB 110918A:} \citet{elliott2013}
\textbf{GRB 111209A:} \citet{stratta2013}
\textbf{GRB 120119A:} \citet{morgan2013}
\textbf{GRB 120815A:} \citet{kruhler2013}
\textbf{GRB 130427A:} \citet{perley2013b}

\newpage\clearpage

\LongTables

\begin{landscape}
\begin{deluxetable}{llcccccccr}
\tablecolumns{10}
\tablewidth{0pt}
\tablecaption{Parent sample}
\tablehead{
\colhead{GRB}          
 & \colhead{$z^{a}$}     
 & \colhead{$T_{\mathrm{0}}^{b}$ [days]}          
 & \colhead{$\beta_{\mathrm{O}}$} 
 & \colhead{$A_{\mathrm{V}}^{\mathrm{H}}$}         
 & \colhead{Ext. law$^{c}$} 
 & \colhead{$T_{90}$ [s]} 
 & \colhead{$E_{\gamma,\mathrm{iso}}$ $[10^{52}$ erg]} 
 & \colhead{Flags$^{d}$}
 & \colhead{Redshift ref.}
 }
\startdata
970508  & 0.835$^{(*)}$   & $<$1, $>$1.8     	 & 0.33 $\pm$ 0.17, 0.90 $\pm$ 0.10   & $\sim$0		     & & 20$^{(1)}$& 0.61 $\pm$ 0.13$^{(2)}$ & & \citet{metzger1997}\\
971214  & 3.418$^{(+)}$   & 0.52			   	 	 & 1.20 $\pm$ 0.13$^{(2)}$                                     & 0.33 $\pm$ 0.08     & C& 35$^{(1)}$& 21.1 $\pm$ 2.4$^{(2)}$  & & \citet{kulkarni1998}\\
980703  & 0.966$^{(+)}$   & 5.3			 	 	 & 0.78$^{(3)}$												   & 0.9 $\pm$ 0.2   	 & C& 412 $\pm$ 9$^{(e)}$& 6.9 $\pm$ 0.8$^{(2)}$  & & \citet{djorgovski1998}\\
990123  & 1.6$^{(*)}$     & 0.92				 	 & 0.65$^{(4)}$												   & $\sim$0			 & & 100$^{(1)}$&239 $\pm$ 28$^(2)$  & R & \citet{kulkarni1999}\\
990510	& 1.619$^{(*)}$   & 					 	 & 0.17 $\pm$ 0.15$^{(5)}$									   & 0.22 $\pm$ 0.07	 & SMC& 75$^{(1)}$&17.8 $\pm$ 2.6$^{(2)}$ & & \citet{vreeswijk2001}\\
991208  & 0.706$^{(*)}$  & 				   	 	 & 0.23 $\pm$ 0.37$^{(6)}$									   & 0.80 $\pm$ 0.29     & MW& $<$60$^{(3)}$&22.3 $\pm$ 1.8$^{(2)}$ & & \citet{castro2001}\\
991216  & 1.02$^{(*)}$   & 1.7				 	     & 0.58 $\pm$ 0.08$^{(7)}$									   & $\sim$0		     & & 15.2$^{(e)}$&67.5 $\pm$ 8.1$^{(2)}$ & & \citet{vreeswijk2006}\\
000301C & 2.0404$^{(*)}$ & 3				 	 	 & 0.70 $\pm$ 0.09$^{(9)}$									   & 0.09 $\pm$ 0.04	 & SMC& 2$^{(4)}$& & & \citet{jensen2001} \\
000418  & 1.1181$^{(+)}$ & 2.5				 	 & 0.81$^{(10)}$											   & 0.96 $\pm$ 0.2		 & C& $<$30$^{(e)}$&9.1 $\pm$ 1.7$^{(2)}$ & & \citet{bloom2003}\\
000911	& 1.0585$^{(+)}$ & 						 & 0.65 $\pm$ 0.40$^{(6)}$ 									   & 0.27 $\pm$ 0.32     & SMC& 500$^{(5)}$&67 $\pm$ 14$^{(2)}$ & & \citet{price2002}\\
000926  & 2.0379$^{(*)}$ & 						 & 1.01 $\pm$ 0.16$^{(6)}$									   & 0.15 $\pm$ 0.07     & SMC& $<$25$^{(6)}$&27 $\pm$ 5.8$^{(2)}$ & & \citet{castroa2003}\\
010222	& 1.477$^{(*)}$  & 					     & 0.76 $\pm$ 0.22$^{(6)}$									   & 0.14 $\pm$ 0.08     & SMC& 280$^{(7)}$&81 $\pm$ 1$^{(2)}$ & & \citet{jha2001}\\
010921  & 0.4509$^{(+)}$ & 						 & 0.81 $\pm$ 1.21$^{(6)}$									   & 0.91 $\pm$ 0.82     & MW& 24.6$^{(1)}$&1.1 $\pm$ 0.11$^{(2)}$ & & \citet{price2002b}\\
011121	& 0.362$^{(*)}$  & 						 & 0.61 $\pm$ 0.13$^{(6)}$									   & 0.39 $\pm$ 0.14     & SMC&$>$10$^{(8)}$&$>$2.7$^{(9)}$ & & \citet{garnavich2003}\\
011211	& 2.140$^{(*)}$  & 0.57					 & 0.56 $\pm$ 0.19$^{(11)}$									   & 0.08 $\pm$ 0.08     & SMC& &6.64 $\pm$ 1.32$^{(2)}$ & & \citet{vreeswijk2006}\\	
020124	& 3.198$^{(*)}$  & 						 & 0.11 $\pm$ 0.80$^{(6)}$									   & 0.28 $\pm$ 0.33     & SMC& 78.6$^{(1)}$&21.5 $\pm$ 7.3$^{(2)}$ & & \citet{hjorth2003}\\	
020813	& 1.225$^{(*)}$  & 1.25				 	 & 0.85 $\pm$ 0.06$^{(12)}$									   & 0.12 $\pm$ 0.04     & C& 90$^{(1)}$&67.7 $\pm$ 10.0$^{(2)}$ & & \citet{barth2003}\\
021004  & 2.3351$^{(*)}$ & 0.35 - 5.5				 & 0.39 $\pm$ 0.12$^{(13)}$									   & 0.26 $\pm$ 0.04	 & SMC& 100$^{(10)}$&4.1 $\pm$ 0.7$^{(2)}$ & R & \citet{moller2002}\\
021211  & 1.006$^{(*)}$  & 0.87					 & 0.69 $\pm$ 0.14$^{(14)}$								   & $\sim$0			 & & 3$^{(1)}$&1.1 $\pm$ 0.13$^{(2)}$ & R & \citet{vreeswijk2006}\\
030226	& 1.986$^{(*)}$  &  						 & 0.57 $\pm$ 0.12$^{(6)}$									   & 0.06 $\pm$ 0.06	 & SMC& 76.8$^{(1)}$&6.7 $\pm$ 1.2$^{(2)}$ & & \citet{klose2004}\\
030227  & 1.39    & 						 & 0.78 $\pm$ 2.17$^{(6)}$									   & 0.38 $\pm$ 1.81     & MW& 15$^{(11)}$& & & \citet{watson2003} \\
030328  & 1.5216$^{(*)}$ & 0.78					 & 0.47 $\pm$ 0.15$^{(15)}$									   & $\sim$0			 & & 140$^{(1)}$&36.1 $\pm$ 4.0$^{(2)}$ & & \citet{maiorano2006}\\
030329  & 0.168$^{(*)}$  & 						 & 0.30 $\pm$ 0.22$^{(6)}$									   & 0.54 $\pm$ 0.22	 & MW& $>$23$^{(12)}$&1.66 $\pm$ 0.2$^{(2)}$ & & \citet{caldwell2003}\\ 
030429	& 2.658$^{(*)}$  & $<$1					 & 0.36 $\pm$ 0.12$^{(16)}$									   & 0.34 $\pm$ 0.04	 & SMC& 10.3$^{(13)}$&1.73 $\pm$ 0.31$^{(2)}$ & & \citet{jakobsson2004b}\\
040924  & 0.858$^{(*)}$  & 						 & 0.63 $\pm$ 0.48$^{(6)}$									   & 0.16 $\pm$ 0.44	 & SMC& 2.39 $\pm$ 0.24$^{(14)}$&0.95 $\pm$ 0.1$^{(2)}$ & N & \citet{wiersema2008}\\
041006	& 0.716$^{(*)}$  & 						 & 0.36 $\pm$ 0.27$^{(6)}$									   & 0.11 $\pm$ 0.33	 & MW& 40 $\pm$ 5$^{(15)}$&8.3 $\pm$ 1.3$^{(2)}$ & & \citet{price2004}\\
041219A & 0.3$^{(!)}$    & $>$0.021				 & $\sim$0.4 $^{(17)}$										   & $\sim$0 			 & & 460 $\pm$ 200$^{(11)}$& & & \citet{gotz2011}\\
050318	& 1.4436$^{(*)}$ & 0.05					 & 1.1 $\pm$ 0.1$^{(18)}$									   & 0.67 $\pm$ 0.35 	 & SMC& 32 $\pm$ 2$^{(17)}$& 1.69 $\pm$ 0.17$^{(16)}$ & B & \citet{berger2005}\\
050319	& 3.240$^{(*)}$  & 0.23					 & 0.98 $\pm$ 0.09$^{(19)}$									   & 0.21 $\pm$ 0.07	 & MW& 151.7$^{(18)}$&4.4 $\pm$ 1.8$^{(19)}$ & B &  \citet{jakobsson2006}\\
050401  & 2.8992$^{(*)}$ & 0.02				 	 & 0.39$^{(20)}$											   & 0.67				 & SMC& 33.3 $\pm$ 2.0$^{(20)}$&40.6 $\pm$ 0.84$^{(16)}$ & B, N & \citet{watson2006}\\
050408	& 1.236$^{(*)}$  & 0.6					 & 0.28 $\pm$ 0.33$^{(21)}$									   & 0.73 $\pm$ 0.18	 & SMC& 34$^{(21)}$&$>$1.3$^{(22)}$ &B & \citet{foley2006}\\
050416A & 0.6528$^{(*)}$ & 0.7					 & 1.14 $\pm$ 0.20$^{(22)}$									   & 0.19 $\pm$ 0.11	 & SMC& 2.4 $\pm$ 0.2$^{(23)}$&0.094 $\pm$ 0.01$^{(16)}$ & B &  \citet{soderberg2007}\\
050502A & 3.793$^{(*)}$  & 0.004 - 0.035			 & 0.8$^{(23)}$												   & $\sim$0			 & & 20$^{(24)}$& & N & \citet{prochaska2005}\\
050525A & 0.606$^{(*)}$  & 0.0029, 0.29	 & 0.60 $\pm$ 0.04, 0.94 $\pm$ 0.10$^{(25)}$  & 0.25 $\pm$ 0.13	 & SMC& 8.8 $\pm$ 0.5$^{(25)}$&2.32 $\pm$ 0.36$^{(16)}$ & B & \citet{foley2005}\\ 
050730  & 3.968$^{(*)}$  & 						 & 0.52 $\pm$ 0.05$^{(5)}$									   & 0.10 $\pm$ 0.10     & SMC& 145.1$^{(18)}$&9.1 $\pm$ 0.4$^{(19)}$ & B,N & \citet{starling2005}\\
050801	& 1.56$^{(!)}$   & $<$0.011				 & 0.85 $\pm$ 0.02$^{(26)}$									   & $\sim$0			 & & 20 $\pm$ 2$^{(26)}$&0.92$^{(26)}$ & & \citet{pasquale2007}\\
050802	& 1.71$^{(*)}$   & 						 & 0.36 $\pm$ 0.26$^{(5)}$									   & 0.21 $\pm$ 0.13     & LMC& 13 $\pm$ 2.0$^{(27)}$&$>$2.0$^{(27)}$ & N & \citet{fynbo2005b} \\
050820A	& 2.6147$^{(*)}$ & 0.025, 0.35			 & 0.57 $\pm$ 0.06, 0.77 $\pm$ 0.08$^{(27)}$				   & $\sim$0			 & & $<$600$^{(18)}$&97.5 $\pm$ 7.7$^{(2)}$ &B & \citet{ledoux2005}\\
050824	& 0.828$^{(*)}$  & 						 & 0.45 $\pm$ 0.18$^{(5)}$									   & 0.14 $\pm$ 0.13     & SMC& 24.8$^{(18)}$&0.15 $\pm$ 0.05$^{(19)}$ &B & \citet{sollerman2007} \\
050904	& 6.295$^{(*)}$  & 1.155					 & 1.2 $\pm$ 0.3$^{(28)}$									   & $\sim$0			 & & 181.7$^{(18)}$&124 $\pm$ 7.7$^{(2)}$ &B & \citet{kawai2006}\\
050922C & 2.1995$^{(*)}$ & 						 & 0.51 $\pm$ 0.05$^{(5)}$  & $\sim$0			 & & 4.54$^{(18)}$&4.53 $\pm$ 0.78$^{(2)}$ & B,N & \citet{jakobsson2006}\\
051109A & 2.346$^{(*)}$  & 						 & 0.42 (fixed) $^{(5)}$									   & 0.09 $\pm$ 0.03	 & SMC& 37.2$^{(18)}$&7.52 $\pm$ 0.88$^{(2)}$ & B,N & \citet{quimby2005} \\
051111	& 1.55$^{(*)}$   & 0.06					 & 0.68 $\pm$ 0.03$^{(29)}$									   & 0.23 $\pm$ 0.07	 & SMC& 64$^{(18)}$&5.75 $\pm$ 1.8$^{(19)}$ & B,N & \citet{hill2005} \\
060124  & 2.296$^{(*)}$  & 						 & 0.57 $\pm$ 0.03$^{(5)}$									   & 0.17 $\pm$ 0.03	 & MW& 658$^{(18)}$&43 $\pm$ 3.4$^{(2)}$ & B & \citet{prochaska2006}\\
060206	& 4.048$^{(*)}$  & 0.0091, 0.066			 & 1.42 $\pm$ 0.58, 0.84 $\pm$ 0.14$^{(30)}$				   & $\sim$0			 & & 7 $\pm$ 2$^{(28)}$&4.10 $\pm$ 0.21$^{(16)}$ & B,N & \citet{fynbo2006}\\
060210	& 3.91$^{(*)}$   & 						 & 0.76 (fixed)$^{(5)}$										   & 1.18 $\pm$ 0.10	 & SMC& 220 $\pm$ 70$^{(29)}$&35.3 $\pm$ 1.9$^{(16)}$ & B,N & \citet{cucchiara2006}\\
060418	& 1.489$^{(*)}$  & 0.0093					 & 0.65 $\pm$ 0.06$^{(31)}$									   & 0.10				 & SMC& 52$^{(18)}$&12.8 $\pm$ 1.0$^{(2)}$ & B,N & \citet{vreeswijk2006b}\\
060502A & 1.5026$^{(*)}$ & 		 				 & 0.71 (fixed)$^{(5)}$										   & 0.50 $\pm$ 0.15	 & SMC& 28.5$^{(18)}$&0.32 $\pm$ 0.1$^{(19)}$ & B & \citet{fynbo2009}\\
060512	& 0.4428$^{(+)}$ & 0.12					 & 0.94 $\pm$ 0.03$^{(33)}$									   & $\sim$0			 & & 11.4$^{(18)}$&0.02 $\pm$ 0.005$^{(19)}$ &B & \citet{bloom2006}\\
060526	& 3.221$^{(*)}$  & 						 & 0.55 $\pm$ 0.20$^{(34)}$									   & 0.06 $\pm$ 0.08	 & SMC& 275.2$^{(18)}$&2.58 $\pm$ 0.26$^{(2)}$ & B,N & \citet{thone2010}\\
060605	& 3.773$^{(*)}$  & 0.07, 0.43				 & 0.54 $\pm$ 0.05, 1.02 $\pm$ 0.05$^{(35)}$				   & $\sim$0			 & & 15$^{(18)}$&2.83 $\pm$ 0.45$^{(2)}$ & B,N & \citet{ferrero2009}\\
060607A	& 3.0749$^{(*)}$ & 						 & 0.72 $\pm$ 0.27$^{(5, 36)}$								   & 0.08 $\pm$ 0.08	 & SMC& 103$^{(18)}$&10.9 $\pm$ 1.55$^{(2)}$ &B & \citet{fox2008}\\
060614	& 0.125$^{(*)}$  & 0.116, $>0.35$			 & 0.30 $\pm$ 0.18, 0.81 $\pm$ 0.08$^{(37)}$				   & 0.05 $\pm$ 0.02	 & SMC& 102 $\pm$ 5$^{(30)}$& 0.25 $\pm$ 0.10$^{(16)}$ &B & \citet{price2006}\\
060729	& 0.5428$^{(*)}$ & 0.57					 & 0.78 $\pm$ 0.03$^{(38)}$									   & 0.07 $\pm$ 0.02	 & SMC& 113$^{(18)}$&0.33 $\pm$ 0.10$^{(19)}$ & B,N & \citet{fynbo2009}\\
060904B & 0.703$^{(*)}$  & 						 & 1.11 $\pm$ 0.10$^{(5)}$									   & 0.08 $\pm$ 0.08	 & SMC& 171.9$^{(18)}$&0.36 $\pm$ 0.07$^{(2)}$ & B,N & \citet{fugazza2006}\\
060908	& 1.884$^{(*)}$  & 0.0093, 0.093			 & 0.33 $\pm$ 0.28$^{(39)}$									   & $\sim$0.09			 & SMC& 18.78$^{(31)}$& 4.41 $\pm$ 0.18$^{(16)}$ & B,R & \citet{fynbo2009}\\
060912A	& 0.937$^{(+)}$  & 0.0081, 0.081			 & 0.6 $\pm$ 0.1$^{(40)}$									   & 0.77 $\pm$ 0.10	 & MW& 5$^{(18)}$&0.79 $\pm$ 0.20$^{(19)}$ & B,N & \citet{jakobsson2006c}\\
060927	& 5.467$^{(*)}$  & 0.066					 & 0.61 $\pm$ 0.05$^{(41)}$									   & 0.21 $\pm$ 0.05	 & SMC& 22.6 $\pm$ 0.3$^{(32)}$& 7.56 $\pm$ 0.46$^{(16)}$ &B &  \citet{ruizvelasco2007}\\
061007	& 1.2622$^{(*)}$ & 0.0035					 & 1.02 $\pm$ 0.05$^{(42)}$									   & 0.48 $\pm$ 0.19	 & SMC& 75 $\pm$ 5$^{(33)}$&101.0 $\pm$ 1.4$^{(16)}$ & B,N & \citet{fynbo2009}\\
061121	& 1.3145$^{(*)}$ &  0.0035, 0.8	 &  0.58 $\pm$ 0.04, 0.67 $\pm$ 0.04$^{(43)}$ 	&0.28 $\pm$ 0.08	 & MW& 81 $\pm$ 5$^{(34)}$& 27.2 $\pm$ 1.8$^{(16)}$ &B & \citet{fynbo2009}\\
061126	& 1.1588$^{(*)}$ & $>$ 0.025				 & $\sim$0.5 $^{(5, 19, 44)}$							 	   	& 0.10 $\pm$ 0.04	 & SMC& 50.3$^{(18)}$& 30 $\pm$ 3$^{(2)}$ & B,R & \citet{perley2008} \\
070125	& 1.547$^{(*)}$  & $>$ 3 					 & 0.59 $\pm$ 0.10$^{(5, 45)}$								   	& 0.11 $\pm$ 0.04	 & SMC& 44$^{(18)}$& 93.0 $\pm$ 9.3$^{(2)}$ &B & \citet{updike2008}\\
070208	& 1.165$^{(*)}$  & 						 & 0.66 (fixed)$^{(5)}$										   	& 0.74 $\pm$ 0.03	 & SMC& 64$^{(18)}$&0.28 $\pm$ 0.15$^{(19)}$ &B & \citet{cucchiara2007}\\
070306	& 1.4959$^{(+)}$ & 1.38					 & 0.7 $\pm$ 0.1$^{(46)}$									   	& 5.45 $\pm$ 0.61	 & SMC& 209$^{(18)}$&6.0 $\pm$ 1.5$^{(19)}$ &B & \citet{jaunsen2008}\\
070419A & 0.9705$^{(*)}$ & 0.035					 & 0.82 (-0.16, +0.08)$^{(47)}$								   	& 0.37 $\pm$ 0.19	 & SMC& 160$^{(18)}$&0.24 $\pm$ 0.10$^{(19)}$ &B & \citet{fynbo2009}\\
070802  & 2.4549$^{(*)}$ & $\sim$ 0.025			 & 0.91 $\pm$ 0.04$^{(48)}$									   	& 1.08 $\pm$ 0.12	 & MW& 16.4$^{(18)}$& &B & \citet{eliasdottir2009}\\ 
071003	& 1.6044$^{(*)}$ & 0.012, 2.7				 & 0.29 $\pm$ 0.49, 0.94 $\pm$ 0.03$^{(49)}$				   	& 0.21 $\pm$ 0.08$^{d}$ & SMC& 148 $\pm$ 1$^{(35)}$&34 $\pm$ 4$^{(35)}$ & B,N & \citet{perley2008b}\\
071010A & 0.985$^{(*)}$  & 0.8					 & 0.76 $\pm$ 0.25$^{(50)}$									   	& 0.62 $\pm$ 0.15	 & SMC& 6 $\pm$ 1$^{(36)}$&0.36 $\pm$ 0.29$^{36}$ & B,N & \citet{covino2008}\\
071020	& 2.1462$^{(*)}$ & 						 & 0.8 (fixed)$^{(5)}$										   	& 0.28 $\pm$ 0.09	 & SMC& 4.2 $\pm$ 0.2$^{(37)}$&8.65 $\pm$ 1.53$^{(16)}$ &B & \citet{fynbo2009}\\
071025	& 4.8$^{(!)}$    & 0.12					 & 0.96 $\pm$ 0.14$^{(51)}$									   	& 1.09 $\pm$ 0.20	 & SN& $>$109$^{(38)}$&65$^{(38)}$ &B & \citet{perley2010}\\
071031	& 2.692$^{(*)}$  & $<$0.05, $>$0.05		 & $\sim$0.85, 0.65$^{(52)}$								   	& $\sim$0			 & &180 $\pm$ 10$^{(39)}$& &B & \citet{fynbo2009}\\ 
071112C & 0.8227$^{(*)}$ & 0.38					 & 0.37 $\pm$ 0.02$^{(38)}$									   	& $\sim$0			 & &15 $\pm$ 12$^{(40)}$ &$>$0.53$^{(40)}$ & B,N & \citet{fynbo2009}\\
080129	& 4.349$^{(*)}$  & 0.043, 0.061,			 & 0.57 $\pm$ 0.27, 0.99 $\pm$ 0.26, 							&  $\sim$0   		& &48$^{(41)}$&$>$6.5$^{(41)}$ & B,N & \citet{greiner2009} \\
		&				   & 1.37, 3.5				 & 1.27 $\pm$ 0.04, 1.52 $\pm$ 0.04$^{(53)}$					& 					& & & & \\
080310	& 2.4274$^{(*)}$ & 						 & 0.42 $\pm$ 0.12$^{(5)}$										& 0.19 $\pm$ 0.05	& SMC& 365 $\pm$ 20$^{(42)}$&3.2 $\pm$ 0.3$^{42}$ &B &  \citet{fynbo2009}\\
080319B	& 0.9382$^{(*)}$ & $<$0.005, 0.005 - 0.023,& $\sim$0.8, 0.4, 0.1, 0.8$^{(54)}$							& 0.07				 & SMC& $>$50$^{(43)}$&142 $\pm$ 3.0$^{(16)}$ & B,R &  \citet{fynbo2009}\\
	    &				   & 0.023 - 0.6, $>$0.6	 &																&  					& & & & \\
080319C & 1.9492$^{(*)}$ & 						 & 0.98 $\pm$ 0.42$^{(5)}$										& 0.59 $\pm$ 0.12	& SMC& 34 $\pm$ 9$^{(44)}$&14.6 $\pm$ 2.6$^{(16)}$ & B,N &  \citet{fynbo2009}\\
080330	& 1.5119$^{(*)}$ & 0.0012 - 0.012, 0.93	 & 0.75 $\pm$ 0.03, 1.05 $\pm$ 0.06$^{(55)}$					& $\sim$0			 & & 67$^{45}$&$<$2.2$^{(45)}$  &B &  \citet{fynbo2009}\\
080413B & 1.1014$^{(*)}$ & $<$0.045, $>$1			 & 0.22 $\pm$ 0.04, 0.90 $\pm$ 0.05$^{(56)}$					& $\sim$0			& & 8 $\pm$ 1$^{(46)}$&1.65 $\pm$ 0.06$^{(16)}$ & B,N &  \citet{fynbo2009}\\
080603A	& 1.6874$^{(*)}$ & 0.17					 & 1.01 $\pm$ 0.05$^{(57)}$										& 0.80 $\pm$ 0.13	& LMC& 150 $\pm$ 2$^{(47)}$&2.2 $\pm$ 0.8$^{47}$ & B,N & \citet{guidorzi2011}\\
080603B & 2.6892$^{(49*)}$ & 0.1					 & 0.53 $\pm$ 0.06$^{(58)}$										& $\sim$0			& & 60 $\pm$ 4$^{(48)}$&9.41 $\pm$ 2.45$^{(16)}$ &B &  \citet{fynbo2009}\\
080607	& 3.0363$^{(*)}$ & 0.0035					 & 1.08	$\pm$ 0.05$^{(59)}$										& 3.07 $\pm$ 0.32	& FM& 79 $\pm$ 5$^{49}$&186 $\pm$ 10$^{(16)}$ &B & \citet{prochaska2009}\\
080710	& 0.8454$^{(*)}$ & $>$0.035				 & 1.00 $\pm$ 0.01$^{(60)}$										& $\sim$0			& & 120 $\pm$ 17$^{(50)}$&0.56$^{(50)}$ & B,N &  \citet{fynbo2009}\\
080721	& 2.5914$^{{*}}$ & 1.725					 & 0.86 $\pm$ 0.01$^{(61)}$										& 0.59 $\pm$ 0.21	& SMC& 16.2 $\pm$ 4.5$^{(51)}$& 121 $\pm$ 10$^{(16)}$ & B,N &  \citet{fynbo2009}\\
080810	& 3.3604$^{(*)}$ & 0.008 - 3.5			 & 0.51 $\pm$ 0.22$^{(62)}$										& $\sim$0			& & 107.7$^{(18)}$&39.1 $\pm$ 0.37$^{(52)}$ & B,N &  \citet{fynbo2009}\\
080913	& 6.7$^{(*)}$	   & 0.078					 & 0.79 $\pm$ 0.03$^{(38)}$										& 0.12 $\pm$ 0.03	& SMC& 8 $\pm$ 1$^{(53)}$&7$^{(53)}$ & B,N & \citet{greiner2009b}\\
080916A & 0.689$^{(*)}$   &					 & 1.20 $\pm$ 0.25$^{(63)}$										& 0.20 (-0.06, +0.25)			& SMC& 63$^{54}$&0.92 $\pm$ 0.03$^{16}$ & B& \citet{fynbo2009}\\
080916C & 4.35$^{(!)}$   & 1.4					 & 0.38 $\pm$ 0.20$^{(64)}$										& $\sim$0			& & 66$^{55}$&6500$^{55}$ & & \citet{greiner2009c}\\
080928	& 1.6919$^{(*)}$ & $>$0.05				 & 1.03 $\pm$ 0.01$^{(65)}$										& 0.12 $\pm$ 0.03	& MW& 233.7$^{(56)}$&1.44 $\pm$ 0.92$^{(56)}$ &B & \citet{fynbo2009}\\	
081007	& 0.5295$^{(*)}$ &				 & 0.86 $\pm$ 0.07$^{(63)}$										& $\sim$0	&  & 12$^{(57)}$&0.17 $\pm$ 0.02$^{(16)}$ &B,R& \citet{jin2013}\\	
081008	& 1.967$^{(*)}$  & 0.017, $\sim$0.3		 & 1.14 $\pm$ 0.05, -0.06 $\pm$ 0.17$^{(66)}$					& 0.46 $\pm$ 0.10  & MW& 185 $\pm$ 35$^{(58)}$&6.3$^{(58)}$ & B & \citet{cucchiara2008a}\\
081029  & 3.8479$^{(*)}$ & $<0.035$, 0.035 - 0.25	 & $\sim$ 0.8, 1.05$^{(67)}$									& $\sim$0			& & 270 $\pm$ 45$^{(59)}$& & B,N & \citet{delia2008}\\
081203A & 2.05$^{(*)}$   &0.008 					 & 0.90 $\pm$ 0.01$^{(68)}$										& 0.08				& SMC& 294 $\pm$ 71$^{(60)}$&35.0 $\pm$ 12.8$^{(16)}$ &B& \citet{kuin2009}\\
081222 & 2.77$^{(*)}$   & 					 & 0.47 (-0.00,+0.12)$^{(63)}$										& $\sim$0				&    & 33$^{(54)}$&25.2 $\pm$ 2.3$^{(16)}$ &B& \citet{cucchiara2008b}\\
090102	& 1.547$^{(*)}$  & 0.1					 & 0.35 $\pm$ 0.08$^{(69)}$										& 0.45 $\pm$ 0.07	& LMC& 27 $\pm$ 2.2$^{(61)}$&21.4 $\pm$ 0.4$^{(16)}$ & B,R& \citet{postigo2009}\\
090313	& 3.3736$^{(*)}$ & $<$0.03, $>$0.1		 & 1.20 $\pm$ 0.05, 1.37 $\pm$ 0.15$^{(70)}$					& $\sim$0			 & & 78 $\pm$ 19$^{(62)}$&$>$3.4$^{62}$ &B & \citet{postigo2010}\\
090323	& 3.568$^{(*)}$  & 						 & 0.65 $\pm$ 0.13$^{(71)}$										& 0.14 $\pm$ 0.03  & SMC& 133.1 $\pm$ 1.4$^{(63)}$&410 $\pm$ 50$^{(64)}$ &B & \citet{cenko2011}\\
090328	& 0.7357$^{(*)}$ & 						 & 1.17 $\pm$ 0.17$^{(5)}$										& 0.18 $\pm$ 0.13  & SMC& 57 $\pm$ 3$^{(63)}$&13 $\pm$ 3.0$^{(64)}$ &B & \citet{cenko2011}\\
090423	& 8.3$^{(*)}$	   & 0.06					 & 0.30 $\pm$ 0.06$^{(72)}$										& $\sim$0		  & & 9.77$^{(18)}$&9.5 $\pm$ 2.0 $^{(18)}$ &B & \citet{tanvir2009}\\
090424	& 0.544$^{(*)}$	   & 					 & 0.55 (-0.06,+0.00)$^{(63)}$										& 1.08 (-0.03,+0.06)  & MW & 49.5$^{(54)}$&3.97 $\pm$ 0.08 $^{(16)}$ &B,R& \citet{jin2013}\\
090618	& 0.54$^{(*)}$   & $>$0.059				 & 0.55 $\pm$ 0.07$^{(74)}$										& 0.25 $\pm$ 0.10 & SMC& 113.3$^{(18)}$&25.3 $\pm$ 2.5$^{(52)}$ &B & \citet{cenko2009c}\\
090709A	& 1.80$^{(+)}$   & 					 & 0.44 (-0.17,+0.00)$^{(63)}$										& 2.37 (-0.35,+0.57) & LMC& 89$^{(54)}$&$<$ 229$^{(16)}$ & & \citet{perley2013a}\\
090902B	& 1.882$^{(*)}$  & 1.9					 & 0.68 $\pm$ 0.11$^{(75)}$										& $\sim$0		  & & 21.9$^{(18)}$&440 $\pm$ 30$^{(52)}$ &B &  \citet{cucchiara2009}\\
090926A & 2.1071$^{(*)}$ & 						 & 0.72 $\pm$ 0.17$^{(5)}$										& 0.13 $\pm$ 0.06 & SMC& 20 $\pm$ 2$^{(18)}$&200 $\pm$ 5$^{(52)}$ & B,N & \citet{delia2010}\\
091018	& 0.971$^{(*)}$  &  $>$0.12				 & 0.58 $\pm$ 0.07$^{(76)}$										& 0.07 $\pm$ 0.02 & SMC& 4.4 $\pm$ 0.6$^{(65)}$&0.80 $\pm$ 0.09$^{(16)}$ & B,N & \citet{chen2009} \\
091029	& 2.752$^{(*)}$  & 0.004-2					 & $\sim 0.40 \pm 0.15^{(77)}$										& $\sim 0$ & & 39$^{(54)}$&8.3$^{(66)}$ & B,N & \citet{chornock2009} \\
091127	& 0.490$^{(*)}$  & 0.06 - 0.65			 & 0.30 $\pm$ 0.02$^{(78)}$									    & 0.11 $\pm$ 0.03 & LMC& 7.1 $\pm$ 0.2$^{(67)}$&1.61 $\pm$ 0.03$^{(16)}$ & B,N & \citet{cucchiara2009b} \\
091208B	&1.063$^{(*)}$  & 					 & 0.94 (-0.08,+0.13)$^{(63)}$									    & 0.40 (-0.20,+0.17) & SMC& 14.8 $^{(54)}$&1.97 $\pm$ 0.06$^{(16)}$ & B,N & \citet{wiersema2009} \\
100219A & 4.6667$^{(*)}$& 0.55					 & 0.60 $\pm$ 0.12$^{(79)}$										& 0.24 $\pm$ 0.06 & SMC& 18.8 $\pm$ 0.5$^{(68)}$&$>$1.4$^{(68)}$ & B,N & \citet{thoene2012}\\
100418A	& 0.6239$^{(*)}$& 0.06 - 1.07, $\sim$7	 & 0.97 $\pm$ 0.08, 1.15 $\pm$ 0.07$^{(80)}$					& 0.16 $\pm$ 0.14 & SMC& 8 $\pm$ 2$^{(69)}$&0.09 $\pm$ 0.03$^{(69)}$ & B,N& \citet{postigo2011a}\\
100621A & 0.542$^{(*)}$ & 0.09					 & 0.79 $\pm$ 0.10$^{(81)}$									    & 3.8 $\pm$ 0.2 	& LMC& 63.6 $\pm$ 1.7$^{(70)}$&4.35 $\pm$ 0.48$^{(16)}$ &B& \citet{jensen2010} \\
100901A & 1.408$^{(*)}$ & 0.06, $>$0.6			 & 0.69 $\pm$ 0.15, 0.81 $\pm$ 0.17$^{(82)}$					& 0.19$\pm$ 0.10	& SMC& 439$^{(71)}$&6.3$^{(71)}$ & B,N & \citet{chornock2010} \\ 
110205A & 2.22$^{(*)}$  & 0.15					 & $\sim$0.5$^{(83)}$											& $\sim$0.2			& SMC& 257 $\pm$ 25$^{(72)}$&43.4 $\pm$ 5.0$^{(72)}$ &  & \citet{cenko2011b}\\
110731A & 2.83$^{(*)}$  & 						 & $\sim$0.66$^{(84)}$											& 0.24 $\pm$ 0.06			& SMC& $\sim 15^{(73)}$&7.6 $\pm$ 0.2$^{(73)}$ & N & \citet{tanvir2011}\\
110918A & 0.984$^{(*)}$  & 	2.25			& 0.70 $\pm$ 0.02$^{(85)}$											& 0.16 $\pm$ 0.06			& SMC& & 190$^{(74)}$ &  & \citet{elliott2013}\\
111209A & 0.677$^{(*)}$  & 	0.008,0.013			& 0.48 $\pm$ 0.02, 0.58 $\pm$ 0.04$^{(86)}$			& 0.24 $\pm$ 0.03			& SMC& & 57 $\pm$ 7$^{(75)}$ &  & \citet{vreeswijk2011}\\
120119A & 1.728$^{(*)}$  & 	 							& 0.92 $\pm$ 0.02$^{(87)}$										& 1.09 $\pm$ 0.16			& FM& & $<$21$^{(76)}$ &  & \citet{morgan2013}\\
120815A & 2.36$^{(*)}$  & 	 	0.085						& 0.78 $\pm$ 0.01$^{(88)}$										& 0.15 $\pm$ 0.02		& SMC& &  &  & \citet{kruhler2013}\\
130427A & 0.34$^{(*)}$  & 	 							&  $\sim 0.7^{(89)}$										& 0.13 $\pm$ 0.06		& LMC & 163 $^{(77)}$& 85$^{(78)}$ &  & \citet{levan2013}
\enddata
\tablecomments{
A collection of GRBs with measured redshift, available afterglow photometry and analysed broadband SEDs: values of host galaxy extinction $A_{\mathrm{V}}^{\mathrm{H}}$ and optical spectral index $\beta_{\mathrm{0}}$ for each case are obtained from the referenced literature. We also report epoch(s) in which the SEDs have been analysed and the form of extinction law used to obtain the value of $A_{\mathrm{V}}^{\mathrm{H}}$. Values in the last four columns (for each GRB) are obtained from the same work, referenced in the fourth column. High energy properties - values of $T_{90}$ and $E_{\gamma,\mathrm{iso}}$ - are given in last two columns.\newline
(a) Redshifts obtained via afterglow spectroscopy (marked *), afterglow photometry (marked !) or host galaxy observations (marked +). The only exception is GRB 030227, for which the redshift was determined via X-ray emission lines.\newline
(b) Times, if known, correspond to the epoch after GRB trigger in which the spectral index is measured.\newline
(c) Acronyms stand for the following extinction parametrisations: $C$ - Milky Way extinction law, given by \cite{cardelli1989}; $FM$ - general Local-Group extinction law given by \cite{fitzpatrick1990}; $MW, LMC, SMC$ - Milky Way, Large Magellanic Cloud and Small Magellanic Cloud extinction laws, given by \cite{pei1992}; $SB$ - starburst extinction law, given by \cite{calzetti1994}; $SN$ - supernova dust extinction law, given by \cite{maiolino2004}.\newline
(d) B - Sample B afterglow, R - reverse-shock candidate, N - no apparent early-time reverse component\newline
(e) The value was obtained from BATSE online catalog: http://www.batse.msfc.nasa.gov/batse/grb/catalog/current/tables/duration$\_$table.txt
\newline
\textbf{Spectral properties: }(1) \cite{galama1998} (2) \cite{wijers1999} 
(3) \cite{bloom1998b} (4) \cite{castro1999b} (5) \cite{kann2010} (6) \cite{kann2006} (7) \cite{garnavich2000} (9) \cite{jensen2001} (10) \cite{klose2000} (11) \cite{jakobsson2003} (12) \cite{covino2003} (13) \cite{holland2003} (14) \cite{holland2004} \cite{pandey2003b} (15) \cite{maiorano2006} (16) \cite{jakobsson2004b} (17) \cite{blake2005} (18) \cite{still2005} (19) \cite{schady2010} (20) \cite{watson2006} (21) \cite{postigo2007} (22) \cite{holland2007} (23) \cite{guidorzi2005}  (25) \cite{blustin2006} (26) \cite{pasquale2007} 
(27) \cite{cenko2006} (28) \cite{tagliaferri2005} (29) \cite{butler2006} (30) \cite{monfardini2006} (31) \cite{molinari2007} (32) \cite{xu2009} (33) \cite{japelj2012} (34) \cite{thone2010} (35) \cite{ferrero2009} (36) \cite{nysewander2009} (37) \cite{mangano2007} (38) \cite{zafar2011} (39) \cite{covino2010} (40) \cite{deng} (41) \cite{ruizvelasco2007} (42) \cite{mundell2007b} (43) \cite{page2007} (44) \cite{gomboc2008} (45) \cite{updike2008} (46) \cite{jaunsen2008} (47) \cite{melandri2009} (48) \cite{kruhler2008} (49) \cite{perley2008} (50) \cite{covino2008} (51) \cite{perley2010} (52) \cite{kruhler2009} (53) \cite{greiner2009} (54) \cite{bloom2009} (55) \cite{guidorzi2009} (56) \cite{filgas2011} (57) \cite{guidorzi2011} (58) \cite{jelinek2012} (59) \cite{perley2011} (60) \cite{kruhler2009b} (61) \cite{starling2009} (62) \cite{page2009} (63) \cite{covino2013} (64) \cite{greiner2009c} (65) \cite{rossi2011} (66) \cite{yuan2010} (67) \cite{nardini2011} (68) \cite{kuin2009} (69) \cite{greiner2011} (70) \cite{melandri2010} (71) \cite{mcbreen2010} (72) \cite{tanvir2009} (73) \cite{cucchiara2011} (74) \cite{cano2011} (75) \cite{pandey2010} (76) \cite{wiersema2012} (77) \cite{filgas2012} (78) \cite{troja2012} (79) \cite{thoene2012} (80) \cite{marshall2011} (81) \cite{kruhler2011} (82) \cite{gomboc2013} (83) \cite{gendre2012}  (84) \cite{ackermann2013} (85) \cite{elliott2013} (86) \cite{stratta2013} (87) \cite{morgan2013} (88) \cite{kruhler2013} (89) \cite{perley2013b}
\newline
\textbf{High energy: }(1) \cite{atteia2003} (2) \cite{ghirlanda2008} (3) \cite{hurley1999} (4) \cite{jensen2001} (5) \cite{price2002} (6) \cite{hurley2000b} (7) \cite{intzand2001} (8) \cite{greiner2003} (9) \cite{garnavich2003} (10) \cite{lazzati2002} (11) \cite{vianello2009} (12) \cite{vanderspek2004} (13) \cite{jakobsson2004b} (14) \cite{donaghy2006} (15) \cite{shirasaki2008} (16) \cite{nava2012} (17) \cite{still2005} (18) \cite{sakamoto2011} (19) \cite{kocevski2008} (20) \cite{sakamoto2005} (21) \cite{sakamoto2005b} (22) \cite{postigo2007} (23) \cite{sakamoto2005c} (24) \cite{guidorzi2005} (25) \cite{dellavalle2006} (26) \cite{pasquale2007} (27) \cite{oates2007} (28) \cite{wozniak2006} (29) \cite{curran2007} (30) \cite{barthelmy2006} (31) \cite{covino2010} (32) \cite{ruizvelasco2007} (33) \cite{schady2007} (34) \cite{fenimore2006} (35) \cite{perley2008b} (36) \cite{covino2008} (37) \cite{tueller2007} (38) \cite{perley2010} (39) \cite{kruhler2009} (40) \cite{huang2012} (41) \cite{greiner2009} (42) \cite{littlejohns2012} (43) \cite{racusin2008} (44) \cite{stamatikos2008} (45) \cite{guidorzi2009} (46) \cite{barthelmy2008} (47) \cite{guidorzi2011} (48) \cite{tueller2008} (49) \cite{stamatikos2008b} (50) \cite{kruhler2009b} (51) \cite{starling2009} (52) \cite{ghirlanda2012} (53) \cite{greiner2009b} (54) \cite{margutti2013} (55) \cite{greiner2009c} (56) \cite{rossi2011} (57) \cite{positigo2011b} (58) \cite{yuan2010} (59) \cite{nardini2011} (60) \cite{ukwatta2008} (61) \cite{gendre2010} (62) \cite{melandri2010} (63) \cite{bissaldi2010} (64) \cite{amati2009} (65) \cite{wiersema2012} (66) \cite{filgas2012} (67) \cite{troja2012} (68) \cite{thoene2012} (69) \cite{marshall2011} (70) \cite{ukwatta2010} (71) \cite{gorbovskoy2012} (72) \cite{cucchiara2011} (73) \cite{ackermann2013} (74) \cite{elliott2013} (75) \cite{stratta2013} (76) \cite{morgan2013} (77) \cite{barthelmy2013} (78) \cite{perley2013b}
}
\label{tab1}
\end{deluxetable}
\clearpage
\end{landscape}

\end{document}